\documentclass[letterpaper,conference]{ieeeconf}
\usepackage[top=53pt, left=54pt, right=54pt, bottom=56pt]{geometry}                                                       % letter

\IEEEoverridecommandlockouts                              % This command is only
\pdfoutput=1
\usepackage{bbm}
\usepackage{mathptmx}
\usepackage{graphicx} % more modern
\usepackage{amssymb}
\usepackage{amsthm,amsfonts,bm}
\usepackage{amsmath,mathtools}
\usepackage{times}

\usepackage{soul}
\usepackage{url}
\usepackage[hidelinks]{hyperref}
\usepackage[small]{caption}
\usepackage{graphicx}
\usepackage{booktabs}
\usepackage{xcolor}

\newtheorem{definition}{Definition}
\newtheorem{lemma}{Lemma}

\newtheorem*{assumption*}{Assumption}
\newtheorem{proposition}{Proposition}
\newtheorem{theorem}{Theorem}

\usepackage{physics}



\usepackage{subfig}

\providecommand{\versionnumber}{05 October, 2023}

\title{\LARGE \bf Stochastic Game with Interactive Information Acquisition:\\
Pipelined Perfect Markov Bayesian Equilibrium\\
\normalsize Version \versionnumber
}

\author{Tao Zhang, Quanyan Zhu% <-this % stops a space
\thanks{}% <-this % stops a space
\thanks{Tao Zhang and Quanyan Zhu are with Department of Electrical and Computer Engineering, New York University, 370 Jay Street, Brooklyn, 11201, NY, USA.
        {\tt\small {\{tz636, qz494\}}@nyu.edu}}%
}

\begin{document}

\maketitle
\thispagestyle{empty}
\pagestyle{empty}

%%%%%%%%%%%%%%%%%%%%%%%%%%%%%%%%%%%%%%%%%%%%%%%%%%%%%%%%%%%%%%%%%%%%%%%%%%%%%%%%
%%%%%%%%%%%%%%%%%%%%%%%%%%%%%%%%%%%%%%%%%%%%%%%%%%%%%%%%%%%%%%%%%%%%%%%%%%%%%%%%

\begin{abstract}
This paper studies a multi-player, general-sum stochastic game characterized by a dual-stage temporal structure per period.
The agents face uncertainty regarding the time-evolving state that is realized at the beginning of each period.
During the first stage, agents engage in information acquisition regarding the unknown state.
Each agent strategically selects from multiple signaling options, each carrying a distinct cost.
The selected signaling rule dispenses private information that determines the type of the agent.
In the second stage, the agents play a Bayesian game by taking actions contingent on their private types.
We introduce an equilibrium concept, Pipelined Perfect Markov Bayesian Equilibrium (PPME), which incorporates the Markov perfect equilibrium and the perfect Bayesian equilibrium.
We propose a novel equilibrium characterization principle termed fixed-point alignment and deliver a set of verifiable necessary and sufficient conditions for any strategy profile to achieve PPME.

% This paper studies a general-sum finite-player stochastic game where each time period is composed of two stages.
% The agents have uncertainty about a payoff-relevant state that arrives at the beginning of each period.
% At the first stage, the agents engage in information acquisition for the unknown state. Each agent selects a signaling rule from multiple options with a cost according to her local interests. 
% The chosen signaling rule sends private information or a type for the agent.
% In the second stage, the agents play a Bayesian game by taking actions contingent on their private types.
% We consider an equilibrium concept, pipelined perfect
% Markov Bayesian equilibrium (PPME), by integrating Markov perfect equilibrium and perfect Bayesian equilibrium.
% We propose an equilibrium characterization principle termed fixed-point alignment and establish a verifiable necessary and sufficient condition for a strategy profile to be a PPME.
\end{abstract}

\IEEEpeerreviewmaketitle
%%%%%%%%%%%%%%%%%%%%%%%%
%%%%%%%%%%%%%%%%%%%%%%%%

%---------------------===================
\section{Introduction}

Driven by advancements in technology and improved computational intelligence, the widespread, cost-effective deployment of sensing systems is making it possible for individual participants in large-scale cyber-physical systems to access and process vast amounts of information for real-time decision-making.
However, this exposition of information concurrently cultivates an era of uncertainty, largely owing to the complex and escalating disparities in information.

Addressing these uncertainties has emerged as a critical cognitive aspect of rational decision-making in environments dominated by asymmetric information from various sources, such as experts and service providers, to mitigate the detrimental effects of uncertainty.
For instance, consider a transportation network featuring multiple heterogeneous traffic information providers (TIPs) (e.g., \cite{wu2021value}).
Here, intelligent vehicles (the agents) subscribe to TIPs to gain insights into global traffic conditions and available routes, thereby reducing information asymmetry.

As rational actors, each intelligent vehicle selects TIP subscriptions by incorporating information accuracy, often derived from customer reviews, and the subscription costs into its anticipated daily travel requirements typically expressed through payoff functions.
Supply-chain service providers (SCPs), for example, may opt for more accurate but costlier TIPs to minimize their expected operational costs due to traffic congestion, whereas individual travelers might tolerate traffic delays and choose less accurate but cheaper TIPs.

The selection of TIPs has direct effects on the routing decisions of intelligent vehicles, which in turn impacts the traffic conditions of the network.
Consequently, these traffic conditions influence the travel costs of all vehicles within the network and trigger information updates from their subscribed TIPs.
A self-interested, rational SCP might even foresee such interactions and strategically choose TIPs and routing decisions to manipulate the network and its competitors.

The heightened emphasis on rationality necessitates a reconsideration of agents' information acquisition from a passive act of receiving information to an integral element of rational behavior.
The significance of this expanded rationality is two-pronged.
Firstly, with the deluge of information that is often irrelevant, deceptive, or even manipulative, agents must possess the ability to identify and gather information that supports their decision-making processes.
Secondly, due to their dynamic interactions with other agents and the environment, the choices of information sources not only influence an agent's local actions but also the decision-making of others, as well as the operation of the environment (through the actions), which in turn affects the agent's own utilities.

A Bayesian agent typically manages uncertainties by relying on priors and forming posterior beliefs \cite{kamenica2011bayesian} about the unobserved elements of the agent, or by establishing belief hierarchies (beliefs about the state as well as others' beliefs; see. e.g., \cite{mathevet2020information,zhang2022bayesian}) in competitive multi-agent environments.
These priors and beliefs fundamentally shape the rational decision-making of agents since they characterize the game's uncertainty.
The concept of Bayesian persuasion studies how a principal can use her informational advantage to strategically reveal noisy information about the state relevant to decision-making to the agents, thereby influencing and manipulating agents' beliefs to induce behavior in her favor \cite{kamenica2011bayesian,celli2020private,babichenko2021bayesian,gan2021bayesian,zhang2022bayesian}.

In this work, we focus on a discrete-time, infinite-horizon stochastic with interactive information acquisition (SGIA) played by a finite group of self-interested agents.
SGIA is structured around two sequential decision-making stages within each time period.
During the initial stage, agents engage in interactive information acquisition, aiming to procure noisy information about an unknown state, which is realized at the beginning of each period.
The nature and quality of the information obtained in this stage subsequently define each agent's type of unique private information.
Each agent $i$ has the autonomy to decide how she gets informed about the unknown state by choosing a specific signaling rule that incurs a cost.
We consider that the choices of signaling rules made by other agents do not directly influence the type generation of agent $i$.
However, they impact agent $i$'s beliefs regarding the unknown state and the types of other agents.
The observation of private types instigates the second stage of the game.
Here, each agent decides on a course of regular action, contingent upon her type.
Subsequently, the state evolves according to a Markovian dynamic, dependent on the current state and the action profiles of the agents.
The decision-making in each stage is simultaneous.

Built upon the concepts of \textit{Markov perfect equilibrium} \cite{maskin2001markov} and \textit{perfect Bayesian equilibrium} \cite{fudenberg2005game}, we propose a new equilibrium notion referred to as the \textit{pipelined perfect Markov Bayesian equilibrium} (PPME).
This concept encapsulates the core consistency between the optimalities of agents' information acquisition and regular action-taking in a Markovian dynamic environment.
%
% each agent's interactive information acquisition and subsequent actions, along with the interactions with other agents in the Markovian dynamic environment.
%

We characterize the PPME based on a principle known as the \textit{fixed-point alignment}.
By fixing the strategies for the information acquisition stage, we first formulate the equilibrium behaviors of the regular action-taking as a constrained optimization problem according to the nonlinear optimization formulation for Nash equilibria of a stochastic game (e.g., \cite{filar1997competitive,prasad2012general,prasad2015two,song2018multi}).
% %
% Proposition \ref{prop:opt} establishes a necessary and sufficient condition for the optimality of the agents' action-taking as part of a PPME given arbitrary fixed signaling rule profiles.
% %
We then propose the \textit{global fixed-point alignment} (GFPA) to characterize the selected signaling rule profiles that match the fixed point of the optimal information acquisition at the first stage to the fixed point from the optimal action choices at the second stage.
The GFPA process can be conceptualized as if there is an information designer who aims to induce certain behaviors (i.e., action-taking) of the agents by designing a set of available signaling rules for the agents to choose.
Involving the agents' autonomy of choosing signaling rules at the first stage distinguishes our model from the equilibrium analyses in existing Bayesian persuasion or information design in static environments (e.g., \cite{kamenica2011bayesian,mathevet2020information,celli2020private,babichenko2021bayesian}) as well as in dynamic models (e.g., \cite{gan2021bayesian,zhang2022bayesian,wu2022sequential}).
By decomposing the problem of GFPA into \textit{local fixed-point alignment} (LFPA) problems, we obtain a set of verifiable conditions known as \textit{local admissibility} by applying a KKT-like process to the LFPA problems.
Under a mild condition, we show that the local admissibility serves as a necessary and sufficient condition, placed on the signaling rules selections and action-takings, for PPME.
Thus, if an algorithm converges to locally admissible points, then it provides a PPME for the stochastic game with interactive information acquisition.

The remainder of the paper is organized as follows.
In Section \ref{sec:PSBG}, In Section \ref{sec:PSBG}, we present a formal description of the stochastic game model and introduce the equilibrium concept of pipelined perfect Markov Bayesian equilibrium (PPME).
Section \ref{subsec:GFPA} introduces the concept of global fixed-point alignment (GFPA), while Section \ref{subsec:GFPA} provides a detailed elaboration on local fixed-point alignment (LFPA). Section \ref{sec:discussion_conclusion} provides discussions and concludes the paper.
Omitted proofs are delegated to the online appendix in \cite{zhang2022forward}.

\begin{figure*}
  \centering
    \includegraphics[width=0.8\textwidth]{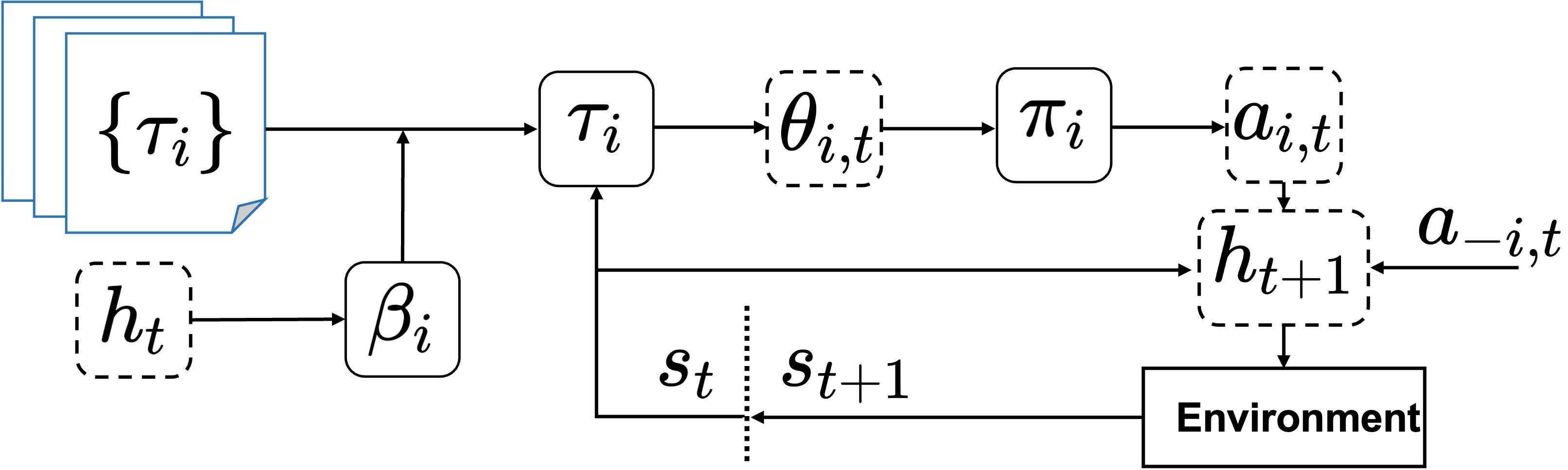}
    \caption{Flow of events in each period of the game $\mathtt{B}^{\Gamma}$. }
    \label{fig:game_play}
\end{figure*}

%%%%%%%%%%%%%%%%%%%%%%%%%
\section{Problem Formulation and Equilibrium}\label{sec:PSBG}

\subsection{Base Game Model}

A finite-player infinite-horizon stochastic game can be characterized by a tuple $\mathtt{B}=\left<N, S, \left\{A_{i}\right\}, \left\{R_{i}\right\}, T, \mathring{T}, \delta\right>$ in which
\begin{itemize}
    \item There is a finite number of players, denoted by $N=\{1,2,\dots, n\}$.
    \item The finite set of states that players can encounter at each period is denoted by $S$.
    \item The finite set of actions that player $i$ can take is denoted by $A_{i}$.
    \item The payoff function of player $i$ is denoted by: $R_{i}:S \times A\mapsto \mathbb{R}$, where $A=\prod_{i\in N} A_{i}$.
    \item The state evolves over time according to $T:S\times A\mapsto \Delta\left(S\right)$; i.e., the probability of $s_{t+1}$ is given by $T(s_{t+1}|s_{t}, a_{t})$ when period-$t$ state is $s_{t}\in S$ and action profile is $a_{t}\in A$. $\mathring{T}(\cdot)\in\Delta(S)$ is the initial distribution of the state. 
    \item $\delta\in(0,1)$ is the common discount factor.
\end{itemize}

For notational compactness, we use \textit{history}, denoted by $h_{t}\in H$, to capture the pair of state and action profile of the last period; i.e., $h_{t}\equiv(s_{t-1}, a_{t-1})$ and $H \equiv S\times A$.
We assume that $h_{t}$ is common knowledge; i.e., $s_{t}$ and $a_{t}$ become publicly observable at the end of each period $t$.

%%%%%%%%%%%%%%%%%%%%%%%%
\subsection{Interactive Information Acquisition}

We consider that the agents do not observe the realizations of the state at the beginning of each period.
Instead, in each period $t$, the agents engage in interactive information acquisition to obtain additional information about the unobserved state $s_{i,t}\in S$.

Based on history $h_{t}\in H$, agents' interactive information acquisition leads to an \textit{information profile}, denote by $\mathcal{I}\left[h_{t}\middle|g_{t}\right]\equiv\left<p(\cdot|h_{t}, g_{t}), \tau(\cdot, g_{t}), \Theta \right>$ at the beginning of each period $t$, where $g_{t}\equiv(g_{i,t})_{i\in N}\in G\equiv \prod_{i\in N}G_{i}$ represents the profile of the agents' choices of information acquisition. 
Here, $\Theta\equiv\prod_{i\in N} \Theta_{i}$ is the profile of finite type spaces for the agents.
$p(\cdot|h_{t}, g_{t})\in\Delta(S\times \Theta)$ is the joint probability of the state and the type profile.
$\tau(\cdot|s_{t}, g_{t})\in \Delta(\Theta)$ is the \textit{signaling rule profile} that generates a type profile $\theta_{t}=(\theta_{i,t})_{i\in N}\in \Theta$ for the agents.
The signaling rule profile is \textit{independent} if $\tau(\theta_{t}|s_{t}, g_{t}) = \prod_{i\in N}\tau_{i}(\theta_{i,t}|s_{t}, g_{i,t})$, or \textit{correlated} if the marginal concerning each agent $i$, $\tau_{i}(\theta_{i,t}|s_{t}, g_{i,t}, g_{-i,t})$, depends on $g_{-i,t}$.
In this work, we focus on independent signaling rules.
In addition, we refer to $g_{t}=(g_{i,t})_{i\in N}\in G$ as the \textit{cognition choice profile} where each $g_{i,t}\in G_{i}$ is agent $i$'s \textit{cognition choice} that indexes the agent's choice of $\tau_{i}(\cdot|s_{t}, g_{i,t})$.
We assume that the cardinalities of $\mathcal{G}_{i}$ and $\Theta_{i}$ are the same for all agents; i.e., $\left|\mathcal{G}_{i}\right|=\left|\mathcal{G}_{j}\right|$ and $\left|\Theta_{i}\right|=\left|\Theta_{j}\right|$, for all $i\neq j$.

Given the state dynamics specified by the base game $\mathtt{B}$, we assume that the information profile $\mathcal{I}\left[h_{t}\middle|g_{t}\right]$ satisfy
\begin{align}
      & \sum\nolimits_{\theta_{t}\in \Theta}p\left(s_{t}, \theta_{t}\middle|h_{t}, g_{t}\right) = T\left(s_{t}\middle|h_{t}\right), \label{eq:common_prior_cond_1}\\
      & \tau\left(\theta_{t}\middle|s_{t}, g_{t}\right) = \frac{p(s_{t}, \theta_{t}|h_{t}, g_{t}) }{T(s_{t}|h_{t})}.\label{eq:common_prior_cond_2}
\end{align}

Given the base game $\mathtt{B}$, define the set of available signaling rule profiles by
\begin{equation}\label{eq:set_of_tau_with_commonprior}
    \mathcal{T}\left[G, \Theta\right]\equiv \left\{\tau\middle| \begin{aligned}
        &\tau_{i}\left(\cdot, g_{i,t}\right):S\mapsto \Delta\left(\Theta_{i}\right), \forall i\in N, \\
        & g_{i,t}\in G_{i}, \textup{s.t. } \exists p(\cdot|h_{t}, g_{t})\in \Delta(S\times \Theta) \\
        & \textup{ satisfying }(\ref{eq:common_prior_cond_1}) \textup{ and } (\ref{eq:common_prior_cond_2}), \forall h_{t}\in H
    \end{aligned} \right\}.
\end{equation}

After all agents made their cognition choices, each agent $i$ privately receives a type $\theta_{i,t}\in \Theta_{i}$ with probability $\tau_{i}(\theta_{i,t}|s_{t}, g_{i,t})$.
Based on his type, agent $i$ chooses an action $a_{i,t}\in A_{i}$.

Each agent $i$'s information acquisition induces a \textit{cognition cost}.
In this work, we restrict attention to the cognition cost that depends on the true state and the action taken by the agent.
That is, after choosing $g_{i,t}$, agent $i$ suffers a cost $C_{i}(s_{t}, a_{i,t})$ that is realized at the end of each period when the true state is $s_{i,t}$ and agent $i$ takes action $a_{i,t}$.
This cost scheme prices agent $i$'s information acquisition based on the consequences of the information acquisition (i.e., the agent's local action $a_{i,t}$) when the true state is $s_{i,t}$.

%%%%%%%%%%%%%%%%%%%%%%%%
\subsection{Stochastic Game with Interactive Information Acquisition}

Let $\Gamma\equiv\left\{\mathcal{T}^{\dagger}\left[G,\Theta\right], C\right\}$ denote the \textit{cognition scheme} for some $\mathcal{T}^{\dagger}\left[G,\Theta\right]\subseteq \mathcal{T}\left[G,\Theta\right]$, where $C=\{C_{i}\}_{i\in N}$.
The base game $\mathtt{B}$ and the cognition scheme $\Gamma$ induces a \textit{stochastic game with interactive information acquisition} (SGIA), denoted by $\mathtt{B}^{\Gamma}$.
Each agent $i$'s decision making in each period $t$ is described as follows.
\begin{itemize}
    \item Contingent on the history $h_{t}$, agent $i$ uses a pure-strategy selection policy $\beta_{i,t}(h_{t})\in G_{i}$ to select $g_{i,t}\in G_{i}$.
    \item Contingent on the type $\theta_{i,t}$, agent $i$ uses a mixed-strategy policy $\pi_{i}$ to choose mixed action $\pi_{i,t}(\theta_{i,t},h_{t})=\left(\pi_{i}(a|\theta_{i,t}, h_{t})\right)_{a\in\mathcal{A}_{i}}\in \Delta(A_{i})$.
\end{itemize}
We say that a policy profile $\pi=(\pi_{i,t}, \pi_{-i,t})_{t\geq 1}$ is \textit{feasible} if it satisfies the following constraints:
%
% \begin{align}
%     &\pi_{i,t}(a_{i,t}|\theta_{i,t})\geq 0, &&\forall a_{i,t}\in A_{i}, \theta_{i,t}\in\Theta_{i}, i\in\mathcal{N}, t\geq 1, \tag{\texttt{FE1}}\label{eq:FE1}\\
%     &\sum\nolimits_{a_{i,t}\in A_{i}} \pi_{i,t}(a_{i,t}|\theta_{i,t})=1, &&\forall \theta_{i,t}\in\Theta_{i}, i\in N, t\geq 1. \tag{\texttt{FE2}}\label{eq:FE2}
% \end{align}
\begin{align}
    \pi_{i,t}(a_{i,t}|\theta_{i,t})\geq 0, \forall a_{i,t}\in A_{i}, \theta_{i,t}\in\Theta_{i}, i\in N, t\geq 1, \tag{\texttt{FE1}}&\label{eq:FE1}\\
    \sum\nolimits_{a_{i,t}\in A_{i}} \pi_{i,t}(a_{i,t}|\theta_{i,t})=1, \forall \theta_{i,t}\in\Theta_{i}, i\in N, t\geq 1. &\tag{\texttt{FE2}}\label{eq:FE2}
\end{align}

With reference to Fig. \ref{fig:game_play}, the following events occur in each period $t$ of $\mathtt{B}^{\Gamma}$.
\begin{itemize}
    \item[1.]  Nature draws a state $s_{t}\in S$ according to $T(\cdot|h_{t})\in\Delta\left( S\right)$.
    \item[2.] Each agent $i$ selects a cognition choice $g_{i,t}\in \mathcal{G}$, based on the common history, which determines $\tau_{i}$. These selections are simultaneous.
    \item[3.] Based on the cognition choice profile $g_{t}$, a type profile $\theta_{t}=\left(\theta_{i,t} \right)_{i\in N}$ is drawn with probability $\tau(\theta_{t}|s_{t}, g_{t})$.
    \item[4.] Each agent $i$ privately observes his type $\theta_{i,t}$ and then chooses an action $a_{i,t}$ with probability $\pi_{i,t}(a_{i,t}|\theta_{i,t})$.
    \item[5.] The state $s_{t}$ and the action profile $a_{t}=(a_{i,t}, a_{-i,t})$ (or, $h_{t+1}=(s_{t}, a_{t})$) become public information, and the state $s_{t}$ is transitioned to a new state $s_{t+1}$ according to $T(\cdot|h_{t+1})\in \Delta(S)$.
\end{itemize}

\subsection{Value Functions}

According to the Ionescu Tulcea theorem \cite{hernandez2012discrete}, $\{\mathring{T}, T, \tau\}$, the agents' policy profile $<\beta, \pi>$, and a cost profile (for a certain cost scheme) uniquely define a probability measure, denoted by $P[\tau,\beta,\pi]$, over $(S\times G \times \Theta \times A)^{\infty}$.
Let $\mathbb{E}^{\tau}_{\beta, \pi}[\cdot]$ denote the expectation operator with respect to $P[\tau,\beta,\pi]$.
In addition, given any $h_{t}$ and $(h_{t}, \theta_{i,t})$, we obtain unique probability measures (perceived by agent $i$) $P[\tau,\beta,\pi|h_{t}]$ and $P[\tau,\beta,\pi|h_{t}, \theta_{i,t}]$ over $S\times G_{-i} \times \Theta \times A \times (S\times G \times \Theta \times A)^{\infty}$ and $S\times \Theta_{-i} \times A\times (S\times G \times \Theta \times A)^{\infty}$, respectively, for all $i\in N$. 
In particular, $P[\tau,\beta,\pi|h_{t}]$ models the uncertainty at period $t$ for each agent $i$ at the beginning of the selection stage while $P[\tau,\beta,\pi|h_{t},\theta_{i,t}]$ models the uncertainty for each agent $i$ at the beginning of the primitive stage.
Let $\mathbb{E}^{\tau}_{\beta, \pi}[\cdot|h_{t}]$ and $\mathbb{E}^{\tau}_{\beta, \pi}[\cdot|h_{t}, \theta_{i,t}]$, respectively, denote the expectation operators with respect to $P[\tau,\beta,\pi|h_{t}]$ and $P[\tau,\beta,\pi|h_{t},\theta_{i,t}]$.

Given $P[\tau,\beta,\pi|h_{t}]$, agent $i$'s period-$t$ \textit{history value function} (H value function) is defined by
\begin{equation}\tag{\texttt{H}}\label{eq:history_value_function}
    \begin{aligned}
    J_{i,t}(h_{t}|\tau, \beta, \pi) \equiv \mathbb{E}^{\tau}_{\beta,\pi}\left[ \sum_{k=t}^{\infty}\delta^{k-t}\left(R_{i}(\tilde{s}_{k}, \tilde{a}_{k}) + C_{i}\left(\tilde{s}_{k}, \tilde{a}_{i,k} \right)\right)\middle| h_{t}\right].
    \end{aligned}
\end{equation}

After a type $\theta_{i,t}$ is realized, each agent $i$ forms a posterior belief, denoted by $\mu_{i}(\cdot|\theta_{i,t}, h_{t})\in \Delta(S \times \Theta_{-i})$, over the state $s_{t}$ and other agents' contemporaneous types $\theta_{-i,t}$.
Due to (\ref{eq:set_of_tau_with_commonprior}), there is a (period-$t$) common prior $p$ such that each posterior belief satisfies
\[
\mu_{i}\left(s_{t}, \theta_{-i,t}\middle|\theta_{i,t}, h_{t}, g_{t}\right) = \frac{p\left(s_{t}, \theta_{i,t}, \theta_{-i,t}\middle| h_{t}, g_{t}\right)}{\sum\nolimits_{s_{t}, \theta_{-i,t} }p\left(s_{t}, \theta_{i,t}, \theta_{-i,t}\middle| h_{t}, g_{t}\right)}.
\]
With abuse of notation, we use $\mu_{i}(s_{t}|\theta_{i,t}, h_{t})$ and $\mu_{i}(\theta_{-i,t}|\theta_{i,t}, h_{t})$ for the marginals of $\mu_{i}(s_{t}, \theta_{-i,t}|\theta_{i,t}, h_{t})$.
Given $h_{t}$ and $\theta_{i,t}$, we define each agent $i$'s expected immediate reward (due to agent $i$'s uncertainty about $s_{t}$) by
\[
\overline{R}_{i}\left(h_{t}, \theta_{i,t}, a_{t}\right) \equiv\sum\limits_{s_{t}\in S } \left(R_{i}(s_{t}, a_{t}) +C_{i,t}\left(s_{t}, a_{i,t} \right)\right)\mu_{i}(s_{t}|\theta_{i,t}, h_{t}).
\]
Given $P[\tau, \beta, \phi|h_{t}]$, agent $i$'s period-$t$ \textit{history-type value function} (HT value function) is defined by
\begin{equation}\tag{\texttt{HT}}\label{eq:HT_value_function}
    \begin{aligned}
    &V_{i,t}(h_{t}, \theta_{t}|\tau, \beta,  \pi)\equiv \mathbb{E}^{\tau}_{\beta,\pi(\cdot|\theta_{t})}\Bigg[\overline{R}_{i}(h_{t}, \theta_{i,t}, \tilde{a}_{t}) \\
    &+\sum_{k=t+1}^{\infty}\delta^{k-t+1}\left(R_{i}(\tilde{s}_{i,k}, \tilde{a}_{k}) + C_{i,k}\left(\tilde{s}_{i,k}, \tilde{a}_{i,k}\right)\right)  \Bigg|h_{t}, \theta_{i,t}\Bigg].
    \end{aligned}
\end{equation}
Here, $V_{i,t}(\cdot)$ depends on $\theta_{-i,t}$ through the policy profile $\pi(\cdot|\theta_{t})$ to take expectation over the current-period action profile.
Finally, agent $i$'s period-$t$ \textit{history-type-action value function} (HTA value function) is defined by
\begin{equation}\tag{\texttt{HTA}}\label{eq:HTA_value_function}
    \begin{aligned}
    &Q_{i,t}(h_{t}, \theta_{i,t}, a_{t}|\tau, \beta,  \pi) \equiv  \overline{R}_{i}(h_{t}, \theta_{i,t}, a_{t})\\
    &+\mathbb{E}^{\tau}_{\beta,\pi}\left[ \sum_{k=t+1}^{\infty}\delta^{k-t+1}\left(R_{i}(\tilde{s}_{i,k}, \tilde{a}_{k}) + C_{i}\left(\tilde{s}_{i,k}, \tilde{a}_{k}\right)\right)  \middle|h_{t}, \theta_{i,t}\right].
    \end{aligned}
\end{equation}
Here, $Q_{i,t}(\cdot)$ is independent of $\theta_{-i,t}$ given the action profile $a_{t}$.

%===============================
\subsection{Pipelined Perfect Markov Bayesian Equilibrium}

In this section, we define a new equilibrium concept for the game $\mathtt{B}^{\Gamma}$.
Our focus lies on the stationary equilibrium, and as such, we omit the time indexes of the value functions and variables, unless explicitly mentioned otherwise.
First, we construct
\begin{align}
    &\mathtt{EV}_{i}\left(h, \theta_{i}\middle|\tau,\beta,\pi; V_{i}\right) \equiv \mathbb{E}^{\tau}_{\beta}\left[V_{i}(h, \theta_{i}, \tilde{\theta}_{-i}|\beta,\tau,\pi)\middle|h, \theta_{i}\right]. \label{eq:EV_i}\\
    & \Pi\left[\beta; \mathtt{B}^{\Gamma}\right]\equiv
    \left\{ \pi\middle| 
     \begin{aligned}
         &\mathtt{EV}_{i}\left(h, \theta_{i}\middle|\tau, \beta, \left(\pi_{i}, \pi_{-i}\right);V_{i}\right)\\
        &\geq \mathtt{EV}_{i}\left(h, \theta_{i}\middle|\tau, \beta, \left(\pi'_{i}, \pi_{-i}\right);V_{i}\right),\\
        &\forall i\in N,  \theta_{i}\in \Theta_{i} ,\pi'_{i}, (\ref{eq:FE1}), (\ref{eq:FE2})
     \end{aligned}
    \right\}.\label{eq:PSR_primitive}
\end{align}

\begin{definition}[Pipelined Perfect Markov Bayesian Equilibrium]
    In a game $\mathtt{B}^{\Gamma}$, a (stationary) strategy profile $\left<\beta, \pi\right>$ constitutes a \textup{pipelined perfect Markov Bayesian equilibrium (PPME)} if for all $i\in N$, $h\in H$, $\theta_{i}\in \Theta_{i}$, it holds in every period that, for $\beta'_{i}$ and $\pi'_{i}$ such that $\left(\pi'_{i}, \pi_{-i}\right)\in \Pi\left[\beta'_{i}, \beta_{-i}; \mathtt{B}^{\Gamma}\right]$,
    \begin{align}
        &J_{i}\left(h \middle|\tau,\left(\beta_{i}, \beta_{-i}\right), \pi\right)  \geq J_{i}\left(h\middle|\tau,\left( \beta'_{i}, \beta_{-i}\right), \left(\pi'_{i}, \pi_{-i}\right) \right),\label{eq:def_ppme_J_1}\\
        & \textit{and } \pi \in \Pi\left[\beta; \mathtt{B}^{\Gamma}\right].
    \end{align}
\end{definition}

The equilibrium concept of PPME builds upon the concepts of Markov perfect equilibrium \cite{maskin2001markov} and perfect Bayesian equilibrium \cite{fudenberg2005game}.

\begin{lemma}\label{lemma:pipeline_subgame_perfect}
    Let $\mathtt{B}^{\Gamma}$ with feasible cognition profile $\Gamma$.
    A strategy profile $\left<\beta, \pi \right>$ is a PPME of a game $\mathtt{B}^{\Gamma}$ if and only if 
    \begin{equation}\label{eq:ppme_refined}
        \begin{aligned}
            &J_{i}\left(h \middle|\tau,\left(\beta_{i}, \beta_{-i}\right), \pi\right)  \geq J_{i}\left(h\middle|\tau,\left( \beta'_{i}, \beta_{-i}\right), \pi \right),\\
            &\textit{and } \pi \in \Pi\left[\beta; \mathtt{B}^{\Gamma}\right].
        \end{aligned}
    \end{equation}
\end{lemma}

Hence, in a PPME problem, each agent $i\in N$ tries to solve the optimization problem given $h\in H$:
\begin{equation}\label{eq:agents_problem}
    \begin{aligned}
        \max\limits_{\beta_{i}, \pi_{i}} &J_{i}\left(h \middle|\tau,\left(\beta_{i}, \beta_{-i}\right), \left(\pi_{i} ,\pi_{-i}\right)\right), 
        \textup{ s.t. } \left(\pi_{i}, \pi_{-i}\right)\in \Pi\left[ \beta_{i}, \beta_{-i}; \mathtt{B}^{\Gamma}\right].
    \end{aligned}
\end{equation}
%
% The decision-making problem (\ref{eq:agents_problem}) belongs to a class of generalized Nash equilibrium problems (GNEP).

\begin{theorem}\label{thm:existence_PPME}
    Fix any $G$ and $\Theta$, there exists at least one profile $\tau\in \mathcal{T}\left[G, \Theta\right]$ such that the game $\mathtt{B}^{\Gamma}$ admits at least one stationary PPME.
\end{theorem}

%%%%%%%%%%%%%%%%%%%%%%%%%%%%%
\section{Equilibrium Characterizations}\label{sec:equilibrium_characterizations}

In this section, we characterize the stationary PPME problem for a given game $\mathtt{B}^{\Gamma}$ by formulating it as a constrained optimization problem and establishing a verifiable condition that is both necessary and sufficient.

Following standard dynamic programming argument (see, e.g., \cite{bellman1966dynamic}), we represent (\ref{eq:history_value_function}), (\ref{eq:HT_value_function}), and (\ref{eq:HTA_value_function}) recursively as follows:
\begin{align}
    &\begin{aligned}
    \mathbf{J}_{i}\left(h\middle|\tau, \beta; V_{i}\right) = \sum_{\theta, s} V_{i}\left(h, \theta\middle|\tau, \beta,\pi\right) \tau\left(\theta\middle|s, \beta\left(h\right)\right)T\left(s\middle|h\right),
    \end{aligned} \label{eq:history_value_function_1}  \\
    &\begin{aligned}
    V_{i}\left(h, \theta\middle|\tau, \beta,\pi\right)= \sum_{a} \pi\left(a \middle|\theta_{i}, \theta_{-i}\right)Q_{i}\left(h, \theta_{i}, a\middle|\tau,\beta,\pi\right),
    \end{aligned}\label{eq:HT_value_function_1} \\
    &\begin{aligned}
    \mathbf{Q}_{i}\left(h, \theta_{i}, a\middle|\tau, \beta;V_{i}\right)&= \overline{R}_{i}\left(h, \theta_{i}, a\right) \\
    &+ \delta \sum\limits_{s}  \mathbf{J}_{i}\left(s, a\middle| \tau, \beta; V_{i}\right) \mu_{i}\left(s\middle|\theta_{i};\tau\right).
    % &Q_{i}(h, g,\theta_{i}, a|\tau;V_{i})= \overline{R}_{i}(h, \theta_{i}, a) + \delta \sum\limits_{s}  J_{i}(s, a, g | \tau; V_{i}) \mu_{i}(s|\theta_{i};\tau).
    \end{aligned}\label{eq:HTA_value_function_1} 
\end{align}
Here, we denote $\mathbf{J}_{i}(\cdot;V_{i})$ and $\mathbf{Q}_{i}(\cdot;V_{i})$ with $V_{i}$ to highlight their dependence on $V_{i}$ from the Bellman recursions (\ref{eq:history_value_function_1})-(\ref{eq:HTA_value_function_1}).
Note that $Q_{i}(\cdot|\beta,\pi,\tau)$ in the right-hand side (RHS) of (\ref{eq:HT_value_function_1}) is given by (\ref{eq:HTA_value_function}).

Leveraging (\ref{eq:history_value_function_1})-(\ref{eq:HTA_value_function_1}), define 
\begin{equation}\tag{\texttt{OBJ1}}\label{eq:obj_1}
    \begin{aligned}
     &Z\left(\pi, V\middle|\beta, \tau\right)\\
     &\equiv \sum_{h, s,\theta }\Bigg(\sum_{i} \left(V_{i}\left(h, \theta\right) - \sum\limits_{a} \mathbf{Q}_{i}\left(h, \theta_{i}, a\middle|\tau,\beta;V_{i}\right)\pi\left(a \middle|\theta\right)\right)\\
     &\times\tau\left(\theta \middle|s, \beta(h) \right)T_{s}\left(s\middle|h\right)\Bigg),
    \end{aligned}
\end{equation}
and in addition, construct the following constraints:
\begin{align}
    &\begin{cases}
    &\begin{aligned}
        &\mathbf{J}_{i}(h|\tau, \beta; V_{i}) \\
        &\geq \sum\limits_{s, \theta_{-i}} V_{i}(h, \theta_{i}, \theta_{-i})\tau_{-i}\left(\theta_{-i}|s,\beta_{-i}(h)\right)T_{s}(s|h),  
    \end{aligned}
       \\
    &\forall i, h,\theta_{i}\textup{ with } \sum_{s}\tau_{i}\left(\theta_{i}|s,\beta_{i}(h)\right) T_{s}(s|h)>0,\tag{\texttt{EQ1}}\label{eq:OB}
    \end{cases}\\
    &\begin{cases}
        &\begin{aligned}
            &\mathtt{EV}_{i}(h, \theta_{i}|\tau,\beta,\pi; V_{i})\\ &\geq \mathbb{E}^{\tau}_{\beta,\pi_{-i}}\Big[ \mathbf{Q}_{i}(h, \theta_{i,t}, a_{i}, \tilde{a}_{-i}|\tau, \beta;V_{i}) \Big|h, \theta_{i}\Big],  
        \end{aligned}\\
    & \forall i, a_{i}, h,\theta_{i}\textup{ with } \sum_{s}\tau_{i}\left(\theta_{i}|s,\beta_{i}(h)\right) T_{s}(s|h)>0. \tag{\texttt{EQ2}}\label{eq:EQ}
    \end{cases}
\end{align}

Define the following set
\begin{equation}\label{eq:set_feasible_pi_V_1}
    \mathcal{K}\left( \beta \middle| \tau\right)\equiv\left\{ 
        \left<\pi,V\right> \Big|\begin{aligned}
             &\pi \textup{ and } V \textup{ satisfy } (\ref{eq:FE1}), (\ref{eq:FE2}), \\
             &(\ref{eq:OB}), (\ref{eq:EQ})
        \end{aligned}
    \right\}.
\end{equation}
Let
\begin{equation}\tag{\texttt{OPT}}\label{set:PPME}
    \mathcal{E}\left( \beta \middle| \tau\right) \equiv \left\{\begin{aligned}
        \arg\min\limits_{\pi, V } &\;  Z(\pi, V|\beta, \tau), \\
        \text{ s.t. } & \left<\pi, V\right>\in  \mathcal{K}\left( \beta \middle| \tau\right)
    \end{aligned}\right\}.
\end{equation}

\begin{proposition}\label{prop:opt}
Fix $\beta$.
In a game $\mathtt{B}^{\Gamma}$, a profile $<\beta, \pi>$ is a PPME if and only if \textit{(i)} $<\pi, V>\in \mathcal{E}\left( \beta \middle| \tau\right)$, where $V=\left(V_{i}\right)_{i\in N}$ is the corresponding optimal \textup{HT} value functions, and \textit{(ii)} $Z(\pi, V|\beta, \tau)=0$. 
\end{proposition}

Proposition \ref{prop:opt} extends the fundamental formulation of finding a Nash equilibrium of a stochastic game as a nonlinear programming (Theorem 3.8.2 of \cite{filar1997competitive}; see also, \cite{prasad2012general,prasad2015two,song2018multi}).
Here, the constraints (\ref{eq:FE1}) and (\ref{eq:FE2}) ensure that each candidate $\pi$ is a valid conditional probability distribution and rules out the possible trivial solution $\{\pi_{i}=0\}_{i\in N}$.
The constraints (\ref{eq:OB}) and (\ref{eq:EQ}) are two necessary conditions for a PPME derived from the optimality of PPME and the Bellman recursions (\ref{eq:history_value_function_1}) and (\ref{eq:HT_value_function_1}).

%====================
\subsection{Global Fixed-Point Alignment}\label{subsec:GFPA}

First, we extend the agents' strategy profile $\left<\beta, \pi\right>$ to $\left<\beta, \pi, V\right>$.
Given $V_{i}$ as a variable, we define the following term based on (\ref{eq:EV_i}):
\begin{equation}\label{eq:EV_i_2}
    \begin{aligned}
    &\mathtt{MV}_{i,t}\left(h_{t}, \theta_{i,t}\middle|\tau,\beta; V_{i,t}\right) \equiv \mathbb{E}^{\tau}_{\beta}\left[V_{i,t}(h_{t}, \theta_{i,t}, \tilde{\theta}_{-i,t})\middle|h, \theta_{i}\right].
    \end{aligned}
\end{equation}
If we fix a $\beta$, Proposition \ref{prop:opt} implies that $\left<\pi, V\right>$ of a PPME profile $\left<\beta, \pi, V\right>$ needs to be a global minimum of (\ref{set:PPME}) with $Z(\pi, V|\tau,\beta)=0$.
Equivalently, given $\pi_{-i}$, each $V_{i}$ needs to be a fixed point of the following equation, for all $i\in N$, $h\in H$, $\theta_{i}\in \Theta_{i}$,
% %
\begin{equation}\tag{\texttt{EQ4}}\label{eq:EQ4}
    \begin{cases}
        &\begin{aligned}
            &\mathtt{MV}_{i}(h,\theta_{i}|\tau,\beta; V_{i})\\
            &\geq \mathbb{E}^{\tau }_{\beta_{-i},\pi_{-i}}\Big[ \mathbf{Q}_{i}(h, \theta_{i}, a_{i}, \tilde{a}_{-i}|\tau,\beta;V_{i}) \Big|\theta_{i}, h\Big],
        \end{aligned}\\
        & \forall a_{i}\in A_{i}, i\in N, h\in H, \theta_{i}\in \Theta_{i}.
    \end{cases}
\end{equation}
% \begin{equation}\tag{\texttt{EQ4}}\label{eq:EQ4}
%     \begin{cases}
%         &\begin{aligned}
%             &\mathtt{MV}_{i}(h,\theta_{i}|\tau,\beta; V_{i})\\
%             &= \max\limits_{a_{i}\in A_{i}} \mathbb{E}^{\tau }_{\beta_{-i},\pi_{-i}}\Big[ \mathbf{Q}_{i}(h, \theta_{i}, a_{i}, \tilde{a}_{-i}|\tau,\beta;V_{i}) \Big|\theta_{i}, h\Big],
%         \end{aligned}\\
%         & \forall i\in N, h\in H, \theta_{i}\in \Theta_{i}.
%     \end{cases}
% \end{equation}
% %
where dependence of the RHS of (\ref{eq:EQ4}) on $V_{i}$ is due to (\ref{eq:HTA_value_function_1}).
At the cognition stage, the optimality of PPME requires that the agents' choice $g$ among all possible options is optimal.
Given any $J_{i}(\cdot)$, define, for all $i\in N$, $\theta_{i}\in\Theta_{i}$, $h\in H$,
\begin{equation}\label{eq:recursive_J_reformulate}
    \begin{aligned}
        &\mathtt{IJ}_{i}(h,\theta_{i}|\tau_{-i},\beta_{-i},\pi;J_{i})\equiv \sum\limits_{s, \theta_{-i},a}\Big( \overline{R}_{i}(h, \theta_{i}, a) + J_{i}\left(s, a\right)\Big) \\
    &\times\pi(a|\theta_{i},\theta_{-i})\tau_{-i}\left(\theta_{-i}|s, \beta_{-i}(h)\right) T(s|h), 
    \end{aligned}
\end{equation}
where $J_{i}(s,a)$ on the RHS of (\ref{eq:recursive_J_reformulate}) is a H value function of the next period given current-period $(s,a)$.
The optimality of $\tau$ in the cognition stage of a PPME (i.e., constraint (\ref{eq:OB})) implies that the optimal history value function $J_{i}$ for each agent $i$ needs to be a fixed point while fixing others' $\tau_{-i}$. That is, 
\begin{equation}\tag{\texttt{EQ3}}\label{eq:OB1}
\begin{cases}
    &J_{i}(h) \geq \mathtt{IJ}_{i}(h, \theta_{i}|\tau_{-i}, \beta_{-i},\pi; J_{i}),\\
    & \forall \theta_{i}\in\Theta_{i}, i\in N, h\in H.
\end{cases}
\end{equation}
Here, (\ref{eq:OB1}) is independent of $V$ while (\ref{eq:EQ4}) is independent of $J$.
In order to make $<\beta, \pi>$ as a PPME of $\mathtt{B}^{\Gamma}$, $<\beta, \pi>$ must be chosen such that there exist $\left<J, V\right>$ satisfying
\[
\left\{\begin{aligned}
    &\textup{$J$ is a fixed point of (\ref{eq:OB1}) if and only if}\\
    &\textup{$V$ is a fixed point of (\ref{eq:EQ4}).}
\end{aligned} \right\}
\]
We refer to such a procedure as the \textit{Global Fixed-Point Alignment} (GFPA).

Since $\beta_{i}$ is a pure strategy, each deterministic choice of $g_{i}=\beta_{i}(h)$ leads to a signaling rule $\tau_{i}(\cdot|\cdot, g_{i})$ that determines a distribution of agent $i$'s period-$t$ types.
That is, every $\beta_{i}$ determines a unique $\tau_{i}(\cdot, g_{i}): S\mapsto \Delta(\Theta_{i})$ for every $h$.
Hence, for ease of exposition, we use $\tau_{i}$ and $\tau_{i,t}(\cdot|\cdot, g_{i}, h)$ (with abuse of notation) to represent $\beta_{i}$ and $\beta_{i}(h)$, respectively; unless otherwise stated.
Therefore, in game $\mathtt{B}^{\Gamma}$, each agent $i$ controls $\left<\tau_{i} ,\pi\right>$.

Given a $\pi$, define the following function of $\tau$, $J$, and $V$:
\begin{equation}\tag{\texttt{OBJ2}}\label{eq:obj_FPA}
    \begin{aligned}
        &Z^{\mathtt{GFPA}}(\tau, J, V|\pi)\\
        &\equiv \sum\nolimits_{i,h}\left(J_{i}(h) - \sum\nolimits_{\theta, s} V_{i}(h, \theta)\tau(\theta|s, g, h)T_{s}(s|h)\right).
    \end{aligned}
\end{equation}
Similar to (\ref{eq:FE1}) and (\ref{eq:FE2}) for $\pi$, we introduce the following two constraints placed on $\tau$:
\begin{align}
        &\tau_{i}(\theta_{i}|s, g_{i},h)\geq0, \forall i\in N, \theta_{i}\in\Theta_{i}, s\in S, g\in G, h\in H,\tag{\texttt{RG1}}\label{eq:regular_tau}\\
        &\sum\nolimits_{\theta_{i}\in \Theta_{i} }\tau(\theta_{i}|s, g_{i}, h) = 1, \forall i\in N,  s\in S, g\in G, h\in H. \tag{\texttt{RG2}}\label{eq:feasible_tau}
\end{align}

Define the following set:
\begin{equation}
    \mathcal{K}^{\mathtt{GFPA}}(\pi)\equiv\left\{ 
        \left<\tau, J, V\right>\middle| 
            (\ref{eq:regular_tau}), (\ref{eq:feasible_tau}), (\ref{eq:OB1}), (\ref{eq:EQ4})
 \right\}.
\end{equation}
Let
\begin{equation}\tag{\texttt{GFPA}}\label{set:FPA}
    \begin{aligned}
      \mathcal{E}^{\mathtt{GFPA}}(\pi) = \left\{\begin{aligned}
          \arg\min\limits_{\tau, J, V}&\; Z^{\mathtt{GFPA}}(\tau, J, V|\pi),\\ \textup{ s.t. }& \left<\tau, J, V\right> \in \mathcal{K}^{\mathtt{GFPA}}(\pi)
      \end{aligned}\right\}.
    \end{aligned}
\end{equation}

\begin{proposition}\label{prop:FPA}
Suppose that $\left<J, V, \pi\right>$ satisfy the Bellman recursions (\ref{eq:history_value_function_1})-(\ref{eq:HTA_value_function_1}).
Then, $\left<\tau, J, V\right>\in\mathcal{E}^{\mathtt{GFPA}}(\pi)$ with $Z^{\mathtt{GFPA}}(\tau, J, V|\pi)=0$ if and only if $\left<\pi, V\right>\in \mathcal{E}\left(\beta\middle|\tau\right)$  with $Z(\pi, V|\tau) = 0$.
\end{proposition}

The proof of Proposition \ref{prop:FPA} is deferred to Appendix \ref{app:prop:FPA}.
Proposition \ref{prop:FPA} shows that if $\left<\tau, J, V\right>\in \mathcal{E}^{\mathtt{GFPA}}(\pi)$ with $Z^{\mathtt{GFPA}}\left(\tau, J, V\middle|\pi\right)=0$ for a policy $\pi$, then the profile $\left<\tau, \pi\right>$ is a PPME.

% \subsection{An Optimization Problem Formulation for PPME}

% Following the simplification
% $\tau_{i}(\cdot)=\tau_{i}(\cdot|\cdot, g_{i}, h)$ to represent the choice of $\tau_{i}$ using $\beta_{i}(h)$, the function $Z(\pi, V|\beta,\tau)$ given by (\ref{eq:obj_1}) can be written as $Z(\pi,V|\tau)$, and $\mathcal{K}(\beta|\tau)$ and $\mathcal{E}(\beta|\tau)$, respectively, given by (\ref{eq:set_feasible_pi_V_1}) and (\ref{set:PPME}) can be written as $\mathcal{E}(\tau)$.
% %
% Thus, a PPME profile $\left<\tau, \pi\right>$ (or $\left<\beta, \pi\right>$) can be obtained by solving
% %
% \begin{equation}
%     \begin{aligned}
%         \min_{\pi,V}&\; Z\left(\pi, V\middle|\tau\right), \\
%         &\textup{ s.t., } \left<\pi, V\right>\in \mathcal{K}\left(\tau\right), \left<\tau, J,V \right>\in \mathcal{E}^{\mathtt{GFPA}}\left(\tau\right).
%     \end{aligned}
% \end{equation}
% %
% Equivalently, a PPME profile $\left<\tau, \pi\right>$ solves 
% %
% \begin{equation}
%     \begin{aligned}
%         \min_{\tau, J, V}&\; Z^{\mathtt{GFPA}}\left(\tau, J, V\middle|\pi \right), \\
%         &\textup{ s.t., } \left<\pi, V\right>\in \mathcal{E}\left(\tau\right), \left<\tau, J,V \right>\in \mathcal{K}^{\mathtt{GFPA}}\left(\tau\right).
%     \end{aligned}
% \end{equation}

% \textcolor{red}{active constraints}

%=======================================
\subsection{Local Fixed-Point Alignment}\label{subsec:LFPA}

First, we decompose each type space $\Theta_{i}$ into $\Theta_{i}=\Theta^{\natural}_{i}\cup \{\hat{\theta}_{i}\}$ such that the constraint  (\ref{eq:feasible_tau}) can reformulated as, for all $i\in N$, $s\in S$, $g\in G$, $h\in H$,  
\begin{equation}%\tag{\texttt{RG3}}\label{eq:feasible_tau_decomp}
    \begin{aligned}
    &\sum\nolimits_{\theta_{i}\in \Theta^{\natural}_{i} }\tau_{i}\left(\theta_{i}\middle|s, g_{i}, h\right) \leq 1,\\ &\tau_{i}\left(\hat{\theta}_{i}\middle|s, g_{i}, h\right)+\sum\nolimits_{\theta_{i}\in \Theta^{\natural}_{i} }\tau_{i}\left(\theta_{i}\middle|s, g_{i}, h\right) = 1.
    \end{aligned}
\end{equation}
Given $\tau_{-i}(\theta_{-i}|s, g_{-i}, h) = \prod_{j\neq i} \tau_{j}(\theta_{i}|s, g_{j}, h)$, define for all $i\in N$,
\begin{equation*}
    \begin{aligned}
    &\mathtt{IV}_{i}(h, \theta_{i}| \tau_{-i};V_{i})\equiv \sum\limits_{\theta_{-i}, s} V_{i}(h, \theta_{i}, \theta_{-i})\tau_{-i}(\theta_{-i}|s, g_{-i}, h)T_{s}(s|h).
    \end{aligned}
\end{equation*}
Construct the vector 
$X^{s}_{i} \equiv \left(J_{i}(h), V_{i}(h,\cdot), \tau_{-i}\left(\cdot|s, g_{-i}, h\right)  \right)$.
Define 
\[
\lambda_{i}\left(X^{s}_{i} ;  \theta_{i}, h  \right) \equiv J_{i}(h) - \mathtt{IV}_{i}(h, \theta_{i}| \tau_{-i};V_{i}).
\]
Here, $\lambda_{i}\left(X^{s}_{i} ;  \theta_{i}, h  \right)$ is a function of $J_{i}$, $V_{i}$, and $\tau_{-i}(\cdot|s,g_{-i}, h)$ and is independent of $\pi$ and $\tau_{i}$.

For any $s\in S$, $h\in H$, $\theta_{i}\in \Theta_{i}$, define the following function
\begin{equation}\label{eq:misalign_def}
    \begin{aligned}
        Z^{\mathtt{LFPA}}_{i}\left(X^{s}_{i}, \tau_{i}; s, h\right)&\equiv \sum\nolimits_{\theta'_{i}\in \Theta^{\natural}_{i} } \lambda_{i}(X^{s}_{i};  \theta'_{i}, h)\tau_{i}(\theta'_{i}|s,g_{i},h) \\
        &+ \lambda_{i}(X^{s}_{i};  \hat{\theta}_{i}, h)\tau_{i}(\hat{\theta}_{i}|s,g_{i}, h).
    \end{aligned}
\end{equation}
Define the following set
\begin{equation}
    \begin{aligned}
        \overline{\mathcal{K}}\left[s,g_{-i}, h\right]\equiv \left\{\left<X^{s}_{i}, \tau_{i},  \right>\middle|\begin{aligned}
            &\tau_{i}(\hat{\theta}_{i}|s, g_{i}, h) \geq 0\\
            & \tau_{i}(\theta_{i}|s, g_{i}, h) \geq 0, \forall \theta_{i}\in \Theta^{\natural}_{i}\\
        &\lambda_{i}(X^{s}_{i}; \theta'_{i}, h)\geq 0, \forall \theta'_{i} \in \Theta_{i}
        \end{aligned}  \right\}.
    \end{aligned}
\end{equation}
Then, we define the problem of \textit{Local Fixed-Point Alignment} (LFPA) by
\begin{equation}\tag{\texttt{LFPA}}\label{eq:LMM}
    \begin{aligned}
        \min_{X^{s}_{i}, \tau_{i}} Z^{\mathtt{LFPA}}_{i}\left(X^{s}_{i}, \tau_{i}; s, h\right), \textup{ s.t. } \left<X^{s}_{i}, \tau_{i} \right>\in \overline{\mathcal{K}}\left[s,g_{-i}, h\right].
    \end{aligned}
\end{equation}
Let $e_{i}$, $\bm{b}_{i}\equiv(b[\theta_{i}])_{\theta_{i}\in \Theta^{\natural}_{i}}$,  $\bm{f}_{i}\equiv(f[\theta_{i}])_{\theta_{i}\in \Theta_{i}}$, respectively, denote the Lagrange multipliers of the constraints $\left\{ \tau_{i}(\hat{\theta}_{i}|s, g_{i}, h) \geq 0\right\}$, $\left\{\tau_{i}(\theta_{i}|s, g_{i}, h) \geq 0, \forall \theta_{i}\in \Theta^{\natural}_{i}\right\}$,  and $\left\{ \lambda_{i}(X^{s}_{i}; \theta'_{i}, h)\geq 0, \forall \theta'_{i} \in \Theta_{i} \right\}$.
In addition, the corresponding slack variables are denoted by $w_{i}$, $\bm{q}_{i}\equiv\{q[\theta_{i}]\}_{\theta_{i}\in \Theta^{\natural}_{i}}$, $\bm{z}_{i}\equiv\{z[\theta_{i}] \}_{\theta_{i}\in \Theta_{i}}$, respectively.
Then, the Lagrangian of (\ref{eq:LMM}) is defined by
\begin{equation}
    \begin{aligned}
        &L_{i}(X^{s}_{i}, \tau_{i}, e_{i}, \bm{b}_{i}, \bm{f}_{i}, w_{i}, \bm{q}_{i},  \bm{z}_{i}|s,h)\equiv Z^{\mathtt{LFPA}}_{i}(X^{s}_{i}, \tau_{i}; s, h) \\
        &+ \sum\nolimits_{\theta_{i}\in \Theta^{\natural}_{i}}b[\theta_{i}]\left(q[\theta_{i}] - \tau_{i}(\theta_{i}|s,g_{i},h) \right)\\
    &+ \sum\nolimits_{\theta_{i}\in \Theta_{i} } f[\theta_{i}]\big( z[\theta_{i}] -  \lambda_{i}(X^{s}_{i}; \theta_{i}, h)\big) + e_{i}\big(w - \tau^{k}_{i}(\hat{\theta}_{i}|s, g_{i}, h)\big). 
    \end{aligned}
\end{equation}
To simplify the presentation, we omit $s$ and $h$.
Taking partial derivatives of $L_{i}$ with respect to $X^{s}_{i}$ and $\tau_{i}$ yields,
\begin{equation*}
    \begin{cases}
         &\begin{aligned}
             &\Delta_{i}\left(X^{s}_{i}, \tau_{i}, \bm{f}_{i}\right)\\
             &\equiv \gradient_{X^{s}_{i}} Z^{\mathtt{LFPA}}_{i}(X^{s}_{i}, \tau_{i}) - \sum_{\theta_{i}\in \Theta_{i}}f[\theta_{i}]\gradient_{X^{s}_{i}}\lambda_{i}(X^{s}_{i}; \theta_{i}),
         \end{aligned}\\
    & \begin{aligned}
        &D_{i}\left(X^{s}_{i}, \tau_{i}(\theta_{i}|\cdot), e_{i}, b[\theta_{i}]\right)\\
    &\equiv b[\theta_{i}] - e + \frac{\partial}{\partial \tau_{i}(\theta_{i}|\cdot)} Z^{\mathtt{LFPA}}_{i}(X^{s}_{i}, \tau_{i}(\theta_{i}|\cdot)), \forall \theta_{i}\in \Theta_{i}.
    \end{aligned}
    \end{cases}
\end{equation*}

Let $\bm{X}^{s}\equiv(X^{s}_{i})_{i\in N }$, $\bm{f}\equiv(\bm{f}_{i})_{i\in N}$, $\bm{e}\equiv(e_{i})_{i\in N}$, $\bm{b}\equiv(\bm{b}_{i})_{i\in N}$, and $\bm{\lambda}\equiv(\lambda_{i})_{i\in N}$.
Define
\[
\begin{cases}
    &\begin{aligned}
        &\bm{F}\left(\bm{X}^{s}, \tau, \bm{e}, \bm{b}, \bm{f}\right)\\
        &\equiv \Big( \Delta_{i}\left(X^{s}_{i}, \tau_{i}, \bm{f}_{i}\right), D_{i}\left(X^{s}_{i}, \tau_{i}(\theta_{i}|\cdot), e_{i}, b[\theta_{i}]\right)\Big)_{i\in N},
    \end{aligned} \\
    &\begin{aligned}
        &\bm{K}\left(\bm{e}, \bm{b}, \bm{f} ;  \tau,  \bm{\lambda}\right)\\
        &\equiv \Big(e_{i}\tau_{i}(\hat{\theta}_{i}), b[\theta_{i}]\tau_{i}(\theta_{i}), f[\theta_{i}]\lambda_{i}(X^{s}_{i};\theta_{i}) \Big)_{i\in\mathcal{N}}.
    \end{aligned} 
\end{cases}
\]
%
% %
% \begin{align*}
%     &\begin{aligned}
%         &\bm{F}\left(\bm{X}^{s}, \tau, \bm{e}, \bm{b}, \bm{f}\right)\\
%         &\equiv \Big( \Delta_{i}\left(X^{s}_{i}, \tau_{i}, \bm{f}_{i}\right), D_{i}\left(X^{s}_{i}, \tau_{i}(\theta_{i}|\cdot), e_{i}, b[\theta_{i}]\right)\Big)_{i\in N},
%     \end{aligned} \\
%     %
%     &\begin{aligned}
%         &\bm{K}\left(\bm{e}, \bm{b}, \bm{f} ;  \tau,  \bm{\lambda}\right)\\
%         &\equiv \Big(e_{i}\tau_{i}(\hat{\theta}_{i}), b[\theta_{i}]\tau_{i}(\theta_{i}), f[\theta_{i}]\lambda_{i}(X^{s}_{i};\theta_{i}) \Big)_{i\in\mathcal{N}}.
%     \end{aligned} 
% \end{align*}
% %
%
Construct the set
\begin{equation}
    \mathcal{R}\left(s,h\right)\equiv \left\{\left<\bm{X}^{s}, \tau, \bm{e}, \bm{b}, \bm{f} \right>\middle|
    \begin{aligned}
        &\bm{F}\left(\bm{X}^{s}, \tau, \bm{e}, \bm{b}, \bm{f}\right)=0\\
        & \bm{K}\left(\bm{e}, \bm{b}, \bm{f} ;  \tau,  \bm{\lambda}\right) = 0
    \end{aligned}
    \right\}.
\end{equation}
For any $\theta_{i}\in \Theta_{i}$, define
\begin{equation}\label{eq:local_S}
    \begin{aligned}
    &\gamma_{i}\left(J_{i}, V_{i}, \pi_{-i}| \tau_{i}, \theta_{i}, a_{i}, h\right) \equiv \mathtt{EV}_{i}\left(h, \theta_{i}|\tau_{i}, \pi, V_{i}\right)\\
    &-\mathbb{E}^{\mu_{i}}_{\pi_{-i}}\left[ Q_{i}(h, \theta_{i}, a_{i}, \tilde{a}_{-i}|\tau; J_{i})\Big|h, \theta_{i} \right],
    \end{aligned}
\end{equation}
where $Q_{i}(\cdot ; J_{i})$ is defined via replacing $J_{i}(\cdot;V_{i})$ in (\ref{eq:HTA_value_function_1}) by $J_{i}(\cdot)$. That is,
\[
Q_{i}\left(h, \theta_{i}, a|\tau;J_{i}\right)= \overline{R}_{i}(h, \theta_{i}, a) + \delta \sum\limits_{s}  J_{i}(s, a) \mu_{i}(s|\theta_{i};\tau).
\]
Define the set
\begin{equation}
    \mathcal{R}^{\dagger}\left(J,V\right)\equiv\left\{ \pi\middle|
    \begin{aligned}
         &\pi_{i}(a_{i}|\theta_{i})\gamma_{i}\left(J_{i}, V_{i}, \pi_{-i}| \tau_{i}, \theta_{i}, a_{i}, h\right) = 0, \\
         & \forall i, a_{i}, \theta_{i}, h, (\ref{eq:FE1}), (\ref{eq:FE2})
    \end{aligned}\right\}.
\end{equation}

We define a set of conditions termed \textit{local admissibility} as follows.
\begin{definition}[Local Admissibility]\label{def:local_admissibility}
    A profile $\left<\tau, \pi, J,V\right>$ is \textup{locally admissible} if $\pi\in \mathcal{R}^{\dagger}\left(J,V\right)$ and $\left<\bm{X}^{s}, \tau, \bm{e}, \bm{b}, \bm{f} \right>\in \mathcal{R}\left(s, h\right)$, for all $s\in S$, $h\in H$.
\end{definition}

\begin{theorem}\label{thm:local_admissible}
    Suppose that $\left\{\gradient_{X^{s}_{i}} \lambda_{i}(X^{s}_{i}; \theta_{i}, h)\right\}_{\theta_{i}\in\Theta_{i}}$ is a set of linearly independent vectors for all $X^{s}_{i}$, $i\in N$, $s\in S$, $h\in H$.
    Then, $\left<\beta^{*}, \pi^{*}\right>$ is a PPME if and only if $<\tau^{*}, \pi^{*}, J^{*}, V^{*}>$ is locally admissible.
\end{theorem}

The proof of Theorem \ref{thm:local_admissible} is deferred to Appendix \ref{app:thm:local_admissible}.
Theorem \ref{thm:local_admissible} provides necessary and sufficient conditions for characterizing the PPME.
In particular, if there is an algorithm converges to a local admissible point $\left<\tau,\pi, J, V \right>$ given a feasible cognition profile $\Gamma$ under a linear independence assumption, then the associated profile $\left<\beta, \pi\right>$ is a PPME.
That is, a locally admissible point $\left<\tau,\pi ,J, V\right>$ achieves $Z^{\mathtt{GFPA}}\left( \tau, J, V\middle|\pi\right)=0$ and $Z\left(\pi, V\middle|\beta, \tau \right)=0$.

%==========================
\subsection{Discussion}\label{sec:discussion_conclusion}

Following the simplification
$\tau_{i}(\cdot)=\tau_{i}(\cdot|\cdot, g_{i}, h)$ to represent the choice of $\tau_{i}$ using $\beta_{i}(h)$, the function $Z(\pi, V|\beta,\tau)$ given by (\ref{eq:obj_1}) can be written as $Z(\pi,V|\tau)$, and $\mathcal{K}(\beta|\tau)$ and $\mathcal{E}(\beta|\tau)$, respectively, given by (\ref{eq:set_feasible_pi_V_1}) and (\ref{set:PPME}) can be written as $\mathcal{K}(\tau)$ and $\mathcal{E}(\tau)$.
With abuse of notation, we additionally rewrite $\mathcal{E}(\tau)$ and $\mathcal{K}^{\mathtt{GFPA}}(\tau)$, respectively, as $\mathcal{E}(\tau,V)$ and $\mathcal{K}^{\mathtt{GFPA}}(\tau, V)$ by fixing an arbitrary $V$.
Hence, $\mathcal{E}(\tau,V)$ and $\mathcal{K}^{\mathtt{GFPA}}(\tau, V)$ becomes sets of profiles $\pi$ and $\left<\tau, J\right>$, respectively.

Suppose that the game $\mathtt{B}^{\Gamma}$ admits at least one PPME.
Then, a PPME profile $\left<\tau, \pi\right>$ (or $\left<\beta, \pi\right>$) of $\mathtt{B}^{\Gamma}$ can be obtained by solving the following bi-level constrained optimization problem:
\begin{equation}\label{eq:opp_1}
    \begin{aligned}
        \min_{\pi,V}&\; Z\left(\pi, V\middle|\tau\right), \\
        &\textup{ s.t., } \left<\pi, V\right>\in \mathcal{K}\left(\tau\right), \left<\tau, J\right>\in \mathcal{E}^{\mathtt{GFPA}}\left(\pi, V\right).
    \end{aligned}
\end{equation}
The problem (\ref{eq:opp_1}) is equivalent to
\begin{equation}\label{eq:opp_2}
    \begin{aligned}
        \min_{\tau, J, V}&\; Z^{\mathtt{GFPA}}\left(\tau, J, V\middle|\pi \right), \\
        &\textup{ s.t., } \pi\in \mathcal{E}\left(\tau,V\right), \left<\tau, J,V \right>\in \mathcal{K}^{\mathtt{GFPA}}\left(\tau\right).
    \end{aligned}
\end{equation}
Let us restrict attention to the problem (\ref{eq:opp_2}).
Proposition \ref{prop:opt} implies that for any fixed $\tau$, any $\pi\in \mathcal{E}\left(\tau,V\right)$ satisfies $Z(\pi,V|\tau)=0$. 
Consider the following set
\begin{equation}
    \mathcal{K}^{\dagger}\equiv \left\{\left<\tau, \pi,V, J\right> \middle|\begin{aligned}
       & Z\left(\pi, V\middle|\tau\right)=0,\\
       & (\ref{eq:FE1}), (\ref{eq:FE2}),(\ref{eq:OB}), (\ref{eq:EQ}),\\
       & (\ref{eq:regular_tau}), (\ref{eq:feasible_tau}), (\ref{eq:OB1}), (\ref{eq:EQ4})
    \end{aligned} \right\}.
\end{equation}
Then, we can reformulate the problem (\ref{eq:opp_2}) as
\begin{equation}\label{eq:opt_final_1}
    \begin{aligned}
        \min_{\tau, J, V} Z^{\mathtt{GFPA}}\left(\tau, J, V\middle|\pi\right), \textup{ s.t., } \left<\tau, \pi, V,J\right>\in \mathcal{K}^{\dagger}.
    \end{aligned}
\end{equation}

The total number of decision variables of the optimization problem (\ref{eq:opt_final_1}) is $\mathtt{NV}= n\times\left|H\right|+n\times\left|H\right|\times\prod_{i\in N}\left|\Theta_{i}\right|+\prod_{i\in N}\left|G_{i}\right|\times\left|H\right|\times\left|\Theta_{i}\right| + \prod_{i\in N}\times\left| \Theta_{i}\right|\times\left|A_{i}\right|$.
Let $\mathcal{K}^{E}$ denote the set of active constraints and the equality constraints.
If we use algorithms that depend on Linear Independence Constraint Qualification (LICQ) to solve (\ref{eq:opt_final_1}), then we require the gradients of $\mathcal{K}^{E}$ be linearly independent.
However, at the global minimum of (\ref{eq:opt_final_1}), the number of active constraints plus the number of equality constraints are at least as great as the number of decision variables; i.e., $\left|\mathcal{K}^{E}\right|\geq \mathtt{NV}$.
In general models of $\mathtt{B}^{\Gamma}$, the linear independence of the gradients of $\mathcal{K}^{E}$ is a relatively restrictive condition.

Theorem \ref{thm:local_admissible} shows that the local admissibility can fully characterize the PPME under a condition that requires $\left\{\gradient_{X^{s}_{i}} \lambda_{i}(X^{s}_{i}; \theta_{i}, h)\right\}_{\theta_{i}\in\Theta_{i}}$ to be a set of linearly independent vectors for every $i\in N$, $h\in H$, and $s\in S$.
For any given $h\in H$ and $s\in S$, $\left|X^{s}_{i}\right|=1+ 2\prod_{j\in N}\left|\Theta_{j}\right| -\left|\Theta_{i}\right|$ for all $i\in N$, which is greater than $\left|\Theta_{i}\right|$.
Consequently, we can assert that the requirement for linear independence amongst $\left\{\gradient_{X^{s}_{i}} \lambda_{i}(X^{s}_{i}; \theta_{i}, h)\right\}_{\theta_{i}\in\Theta_{i}}$ is generally less restrictive compared to the necessity for linear independence among the gradients of $\mathcal{K}^{E}$.

The local admissibility (Definition \ref{def:local_admissibility}) can be decomposed into two parts.
First, $\pi\in \mathcal{R}^{\dagger}\left(J,V\right)$ specifies conditions for a policy profile $\pi$ given $J$ and $V$.
Second, $\left<\bm{X}^{s}, \tau, \bm{e}, \bm{b}, \bm{f} \right>\in \mathcal{R}\left(s, h\right)$, for all $s\in S$, $h\in H$, which is independent of the profile $\pi$.
The second part of the local admissibility implies that any algorithm that searches for the zero of the gradients of the Lagrangian of (\ref{eq:LMM}), while converging to $\pi\in \mathcal{R}^{\dagger}\left(J,V\right)$, converges to a PPME.
Designing algorithms that converge to the local admissibility will be our future work.

%============================
\section{Perfect Information Cognition Choice}

In this section, we study when the set of available signaling rule profiles contains a signaling rule that releases the true realizations of the states in every period.

%================
\subsection{Cognition Cost Schemes}

Each agent $i$ chooses a cognition choice $g_{i,t}$ with a cost $C_{i}\in \mathbb{R}$. 
Define $\mathcal{U}_{i}\equiv\left\{G_{i}, S, \Theta_{i},  A_{i}, \Delta\left(S \right), \Delta\left(\Theta_{i}\right)\right\}$.
Let $P'\left(\mathcal{U}_{i}\right)$ denote the power set without the empty set that includes all non-empty subsets of $\mathcal{U}_{i}$.
Then, we define the set of cost functions $C=\left(C_{i}\right)_{i\in\mathcal{N}}$ that can have as their domain any non-empty combination of $G_{i}, S, \Theta_{i}, A_{i}$, $\Delta\left(S \right)$, and $\Delta\left(\Theta_{i}\right)$ by
\[
\mathcal{F}\equiv\left\{C=\left(C_{i}\right)_{i\in N}\middle| C_{i}: X\mapsto \mathbb{R}, \textup{ for } X\in P'\left(\mathcal{U}_{i}\right), \forall i\in N\right\}.
\]

Here are some examples of cognition functions.
The cognition function $C\in \mathcal{F}$ is \textit{cognition-based} (\texttt{CB}) if each $C_{i}: G_{i}\mapsto \mathbb{R}$; that is, $C_{i}(g_{i,t})\in \mathbb{R}$ is the cost if agent $i$ chooses $g_{i,t}\in G_{i}$.
The \texttt{CB} cost directly prices each agent $i$'s cognition choice.
The cost function $C\in \mathcal{F}$ is \textit{type-based} (\texttt{TB}) if each $C_{i}: \Theta_{i}\mapsto \mathbb{R}$; that is, agent $i$ suffers a cost $C_{i}(\theta_{i,t})$ if a type $\theta_{i,t}$ is realized to him according to his cognition choice.
With the \texttt{TB} cost, each agent $i$'s cognition decision is priced based on the realized type.
The cost function $C\in\mathcal{F}$ is \textit{state-type-based} (\texttt{STB}) if each $C_{i}: S\times \Theta_{i}\mapsto \mathbb{R}$; that is, each agent $i$'s cognition cost is $C_{i}(s_{t}, \theta_{i,t})$ if the state is $s_{t}$ and agent $i$ receives a type $\theta_{i,t}$.
The \texttt{STB} cost takes into account the realized state and each agent's type. This cost scheme can capture the settings when the pricing of cognition depends on the difference between the information (about the state) encapsulated in the type and the state.
The cost function $C\in\mathcal{F}$ is \textit{state-action-based} (\texttt{SAB}) if each $C_{i}: S\times \mathcal{A}_{i}\mapsto \mathbb{R}$. That is, each agent $i$ suffers a cost $C_{i}(s_{t}, a_{i,t})$ that depends on the true state $s_{t}$ and his local action $a_{i,t}$. This cost scheme prices agent $i$'s cognition based on the consequences of the information acquisition (i.e., $a_{i,t}$) given the true state.
The cost function $C\in\mathcal{F}$ is \textit{mutual information} (\textit{MI}) if each $C_{i}(\cdot):\Delta\left(\Theta_{i}\right)\times \Delta\left( S\right)\mapsto \mathbb{R}$ is defined by
\[
\begin{aligned}
C_{i}\left(\tilde{\theta}_{i,t};\tilde{s}_{t} \right)\equiv H\left(\tilde{\theta}_{i,t}\right) - H\left(\tilde{\theta}_{i,t}\middle|\tilde{s}_{t}\right),
\end{aligned}
\]
where $H(\cdot)$ denotes the conditional entropy operator (see, e.g., \cite{cover1999elements}).

Let $\Gamma\equiv\left\{\mathcal{T}\left[G,\Theta\right], C\right\}$ denote the \textit{cognition profile} for some $C\in \mathcal{F}$.
We assume that the cardinalities of $G_{i}$ and $\Theta_{i}$ are the same for all agents; i.e., $\left|G_{i}\right|=\left|G_{j}\right|$ and $\left|\Theta_{i}\right|=\left|\Theta_{j}\right|$, for all $i\neq j$.

\subsection{Perfect-Information PPME}

In this section, we introduce PPME under perfect information. 
We start by focusing on a general non-stationary case.

\begin{definition}[Perfect Information Structure]
    An information structure $\{\tau, \Theta\}$ is perfect-information if $\Theta_{i,t} = S$ and there exists $g^{*}_{i}\in G_{i}, \tau_{i,t}(s|s, g^{*}_{i}, h)=1$ for all $i\in N, s\in S, h\in  H$.
% %
% \hfill $\triangle$
%     %
\end{definition}
We use $\{\xi,  S\}$ to denote the perfect(-information) information structure, where $\xi_{i}(s)=s$ for all $s\in S$, $i\in N$.
Let $\mathcal{T}\left[\xi\right]\equiv\left[\xi; G, \Theta \right]$ with $g^{*}_{i}\in G_{i}$ and $\Theta^{g^{*}}_{i}= S$ denote the menu of signaling rules that contains $\left\{\xi,  S\right\}$, and let $\Gamma[\xi]=\left\{T\left[\xi;G, \Theta\right], \left\{C, C\right\} \right\}$ denote the corresponding cognition profile.

With abuse of notation, we use $g^{*}_{i,t}(=g^{*}_{i})\in G_{i}$ to denote agent $i$'s cognition choice that leads to perfect information structure.
Suppose that all other agents choose $\{\xi, S\}$ in every period $t$ in the cognition stage.
Hence, each agent $i$ knows that other agents observe the true $s_{t}$ (i.e., $\theta_{-i,t}= s_{t}$) though agent $i$ may not observe $s_{t}$ (i.e., $g_{i,t}\neq g^{*}_{i,t}$).
For simplicity, we use $\tau_{i,t}$ as agent $i$'s signaling rule due to his period-$t$ cognition choice $g_{i,t}$ without specifying $g_{i,t}$; the same simplification is made for $\xi$ and $g^{*}$.
Hence, agent $i$'s value functions (\ref{eq:history_value_function})-(\ref{eq:HTA_value_function}) can be rewritten as
\begin{align}
    &\begin{aligned}
        &J_{i,t}\left(h_{t} \middle|\tau_{i,t}, \xi_{-(i,t)}, \pi\right) \\
        &= \mathbb{E}^{\tau_{i,t},\xi_{-(i,t)}}_{\pi}\left[ \sum_{k=t}^{\infty}\delta^{k-t}\left(R_{i}(\tilde{s}_{i,k}, \tilde{a}_{k}) + \tilde{c}_{i,k}\right)\middle| h_{t} \right], 
    \end{aligned} \tag{\texttt{PI-H}}\label{eq:PI_history_value_function}\\
     &\begin{aligned}
        &V_{i,t}\left(h_{t}, \theta_{i,t}\middle|\tau_{i,t}, \xi_{-(i,t)},  \pi\right)= \mathbb{E}^{\tau_{i}, \xi_{-(i,t)}}_{\pi}\Big[\overline{R}_{i}(h_{t}, \theta_{i,t}, \tilde{a}_{t}) + \tilde{c}_{i,t} \\
        &+\sum_{k=t+1}^{\infty}\delta^{k-t}\left(R_{i}(\tilde{s}_{i,k}, \tilde{a}_{k}) + \tilde{c}_{i,k}\right)  \Big|h_{t}, \theta_{i,t}\Big],
    \end{aligned} \tag{\texttt{PI-HT}}\label{eq:PI_HT_value_function} \\
     &\begin{aligned}
       & Q_{i,t}\left(h_{t}, \theta_{i,t}, a_{t}\middle|\tau_{i}, \xi_{-(i,t)},  \pi\right) =  \overline{R}_{i}(h_{t}, \theta_{i,t}, a_{t})+c_{i,t} \\
        &+\mathbb{E}^{\xi_{-(i,t)}}_{\pi}\left[ \sum_{k=t+1}^{\infty}\delta^{k-t}\left(R_{i}(\tilde{s}_{i,k}, \tilde{a}_{k}) + \tilde{c}_{i,k}\right)  \middle|h_{t}, \theta_{i,t}\right].
    \end{aligned}\tag{\texttt{PI-HTA}}\label{eq:PI_HTA_value_function}
\end{align}
Define the set
\begin{equation}
    \begin{aligned}
        &\Pi\left[\xi; \mathtt{B}^{\Gamma\left[ \xi\right]}\right]\\
        &\equiv\left\{\pi=(\pi_{i,t})\middle|\begin{aligned}
        &V_{i,t}\left(h_{t}, s_{t}\middle| \xi, \pi\right) \geq V_{i,t}\left(h_{t}, s_{t}\middle| \xi, \left(\pi_{i,t}, \pi_{-(i,t)}\right)\right),\\
        & \forall i\in N, t\geq 1, h_{t}\in H, s_{t}\in S, (\ref{eq:FE1}), (\ref{eq:FE2})
    \end{aligned} \right\}.
    \end{aligned}
\end{equation}

The perfect-information PPME (PI-PPME) is defined as follows.

\begin{definition}[\textup{PI-PPME}]
    In a game $\mathtt{B}^{\Gamma\left[\xi\right]}$, a profile $\left<\xi, \pi\right>$ constitutes a \textup{PI-PPME} if for all $i\in N$, $h\in  H$, it holds in every period $t$ that, for all $i\in N$, $t\geq 1$, $h_{t}\in  H$, $\tau_{i,t}\in \mathcal{T}\left[\xi;  G, \Theta \right]$,
    \begin{equation}
        \begin{aligned}
            &J_{i,t}\left(h_{t}\middle|\xi, \pi\right)\geq J_{i,t}\left(h_{t}\middle|\tau_{i,t}, \xi_{-(i,t)},\pi\right),\\
            &\textit{ and } \pi\in\Pi\left[\xi; \mathtt{B}^{\Gamma\left[ \xi\right]}\right].
        \end{aligned}
    \end{equation}
\end{definition}

For any profile $\left<\tau, \pi\right>$, define each agent $i$'s period-$t$ \textit{ex-post} \textit{history-state value function} (EP-HSA value function) by 
\begin{equation}\tag{\texttt{HSA}}\label{eq:HSA_value_function}
    \begin{aligned}
        &W_{i,t}(s_{t}, a_{t}|\tau,  \pi)\equiv R_{i}(s_{t}, a_{t})+c_{i,t} \\
        &+ \mathbb{E}^{\tau}_{\pi}\left[ \sum_{k=t+1}^{\infty}\delta^{k-t}\left(R_{i}(\tilde{s}_{i,k}, \tilde{a}_{k}) + \tilde{c}_{i,k}\right)  \middle|s_{t}, a_{t}\right].
    \end{aligned}
\end{equation}

Theorem \ref{thm:equivalence_PI_PPME} establishes a relationship between a PPME and a PI-PPME in terms of the H value functions.
\begin{theorem}\label{thm:equivalence_PI_PPME}
    Fix a base game $\mathtt{B}$.
    For a profile $\left<\tau, \pi \right>$ that constitutes a PPME of a game $\mathtt{B}^{\Gamma}$ with feasible cognition profile $\Gamma=\left\{\mathcal{T}^{\natural}\left[G', \Theta'\right], C \right\}$ for any $C\in\mathcal{F}$, there exists a PI-PPME profile $\left<\xi, \pi^{*}\right>$ of a game $\mathtt{B}^{\Gamma\left[\xi\right]}$ with feasible cognition profile $\Gamma\left[\xi\right] = \left\{\mathcal{T}^{\natural}\left[\xi; G, \Theta\right], C^{*}  \right\}$ for \textup{\texttt{SAB}} cost profile  $C^{*}$, such that, for all $i\in N$, $t\geq 1$,$h_{t}\in H$,
    \begin{equation}
        \begin{aligned}
            &J_{i,t}(h_{t}|\tau, \pi) = \mathbb{E}^{\tau}_{\pi}\left[ W_{i,t}\left(\tilde{s}_{t}, \tilde{a}_{t} \middle|\xi, \pi^{*}\right) \middle|h_{t}\right]\\
            &=\mathbb{E}^{\xi}_{\pi^{*}}\left[ W_{i,t}\left(\tilde{s}_{t}, \tilde{a}_{t} \middle|\xi, \pi^{*}\right) \middle|h_{t}\right]=J_{i,t}(h_{t}|\xi, \pi^{*}).
        \end{aligned}
    \end{equation}
\end{theorem}

%%%%%%%%%%%%%%%%
\subsection{Value-Preserving Transformation: From PI-PPME to PPME}

Given any history $h\in H$, we define the \textit{perfect-information \textup{(PI)} achievable value set}, $\mathcal{V}\left(h\right)$ as follows:
\begin{equation}
    \begin{aligned}
        &\mathcal{V}\left(h\right)\equiv \Big\{ u=(u_{i})\Big| u_{i} = J_{i}\left(h\middle|\xi, \pi^{*}\right), \\
        &\textup{ for a PI-PPME $\left<\xi, \pi^{*}\right>$ with a \texttt{SAB} cost profile}  \Big\}.
    \end{aligned}
\end{equation}
For any PI-PPME with a \texttt{SAB} cost profile, let $\mathcal{V}^{*}(h) = \mathcal{V}^{*}(h|\xi,\pi^{*})\equiv\left\{u\middle| u_{i}=J_{i}(h|\xi,\pi^{*}) \right\}=\subset \mathcal{V}\left(h\right)$.
For any $J(h)=\left(J_{i}\left(h\right)\right)_{i\in N}$ in $\mathcal{V}^{*}\left(h\right)$, there must exist a PI-PPME with a \texttt{SAB} cost profile that leads to H value $J_{i}(h)$ for each agent $i$ given history $h\in H$.
Construct
\[
\begin{aligned}
    &\mathbf{V}_{i}\left(h, \theta_{i}\middle|\tau, \pi; J^{*} \right) = \sum_{s,a ,\theta_{-i}} \Big(R_{i}(s, a) + c_{i,t} \\
    &+\delta J^{*}_{i}(s,a) \Big)\pi(a|\theta_{i}, \theta_{-i})\mu_{i}(s, \theta_{-i}|\theta;\tau)T(s|h).
\end{aligned}
\]
Define the function $U: \prod\limits_{h\in\mathcal{H}} \left(\mathcal{V}^{*}(h)\times\prod\limits_{i\in\mathcal{N}}\Delta\left(\Theta_{i}\right)\right)\mapsto\prod\limits_{i\in N}\prod\limits_{h\in H} \Delta\left(\Theta_{i}\right)$ by
\begin{equation}
    \begin{aligned}
        &U\left(\tau\right)(\theta_{i},h|\pi;J(h)) \\
        &= \frac{\tau_{i}\left(\theta_{i}\middle|h\right) +\max\left(0, \mathbf{V}_{i}\left(h, \theta_{i}\middle|\tau,\pi; J^{*} \right)- J^{*}(h)\right) }{1 + \sum_{\theta'_{i}\in \Theta_{i}} \max\left(0, \mathbf{V}_{i}\left(h, \theta'_{i}\middle|\tau,\pi; J^{*} \right)- J^{*}(h)\right) }.
    \end{aligned}
\end{equation}

\begin{proposition}
     Given any PI-PPME profile $\left<\xi, \pi^{*} \right>$ with a \texttt{SAB} cost profile, there exists at least profile $\left<\tau, \pi\right>$ with with a feasible cognition profile with any cost profile $C\in\mathcal{F}$ such that 
     \begin{equation}\tag{\texttt{fp}}\label{eq:fixed_point_tau_PIPPME}
        \tau_{i}\left(\theta_{i}\middle|h\right) = U\left(\tau\right)(\theta_{i},h|\pi;J(h)).
    \end{equation}
\end{proposition}

\begin{theorem}
    Given any PI-PPME profile $\left<\xi, \pi^{*} \right>$ with a \texttt{SAB} cost profile, a profile $\left<\tau,\pi\right>$ with a feasible cognition profile with any cost profile $C\in\mathcal{F}$ is a PPME if and only if, it satisfies (\ref{eq:fixed_point_tau_PIPPME}).
\end{theorem}

% \textcolor{blue}{Algorithm tracks local admissible points}

% \textcolor{blue}{
% %
% Any algorithm that tracks the zero of the gradient of XXX converges to locally admissible points can provide a PPME, in which $\pi$ is updated to ensure convergence to $\mathcal{R}^{\dagger}(J,V)$.
% }

% \textcolor{red}{$$AAAAAA$$}

% % \section{Conclusion}\label{sec:conclusion}

% In this paper, we have studied a dynamic Bayesian persuasion defined as forward-looking dynamic persuasion (FLDP) in a novel pipelined stochastic Bayesian game (PSBG) in which each agent periodically makes a pipeline decision of selecting an information source (IS) and then taking an action based on the information acquired from the selected IS.
% A principal who controls one of the ISs aims to conduct FLDP by first incentivizing the agents to select her IS and then take actions that coincide with her goal.
% We have proposed a design principle termed fixed-point alignment that formulates the principal's design as a non-linear optimization problem and characterized a set of verifiable necessary and sufficient conditions defined as local admissibility for the implementability of the design in a refinement of Nash equilibrium referred to as the pipelined perfect Markov Bayesian equilibrium.
% %
% %
% Designing algorithms that converge to the local admissibility will be our future work.

% %----------------------------------------
%\newpage
% {\footnotesize
% \printbibliography
% }
\bibliographystyle{IEEEtran}
\bibliography{IEEEabrv,TAO_LCSL}

\appendix

\subsection{ Proof of Lemma \ref{lemma:pipeline_subgame_perfect}}\label{app:lemma:pipeline_subgame_perfect}

    The \textit{only if} part is straightforward. In particular, if $\left<\beta, \pi\right>$ is a PPME, then $\pi=(\pi_{i}, \pi_{-i})\in \Pi\left[\beta; \mathtt{B}^{\Gamma}\right]$ and (\ref{eq:def_ppme_J_1}) holds for all $\pi'_{i}$. 
    Hence, (\ref{eq:def_ppme_J_1}) also holds for $\pi_{i}$. 

    We proceed with the proof of the \textit{if} part by establishing a contradiction.
    Suppose that there exists a pair $(\beta'_{i,t}, \pi'_{i,t})$ such that 
\begin{equation}\label{eq:proof_subperfect_1}
        J_{i}\left(h\middle|\tau, \beta, \pi\right)< J_{i}\left(h\middle|\tau, \left(\beta'_{i,t}, \beta_{-(i,t)} \right), \left(\pi'_{i,t},  \pi_{-(i,t)}\right)  \right).
    \end{equation}

The H value function $J_{i}$ can be constructed in terms of $\mathtt{EV}_{i}$ as follows:
   \[
   J_{i}\left(h\middle|\beta, \pi\right) = \sum_{\theta_{i},s} \mathtt{EV}_{i}\left(h,\theta_{i}\middle|\tau, \beta, \pi; V_{i}\right)\tau_{i}\left( \middle|s, \beta_{i}(h)\right)T(s|h).
   \]
Since $\pi \in \Pi\left[\beta; \mathtt{B}^{\Gamma}\right]$, we obtain
 \[
   \begin{aligned}
       &\mathtt{EV}_{i}\left(h, \theta_{i}\middle|\tau, \beta, \pi;V_{i}\right)=\mathtt{EV}_{i}\left(h, \theta_{i}\middle|\tau, \left(\beta'_{i,t}, \beta_{-(i,t)}\right), \pi;V_{i}\right)\\
       &\geq \mathtt{EV}_{i}\left(h, \theta_{i}\middle|\tau, \beta, \left(\pi'_{i,t}, \pi_{-(i,t)}\right);V_{i}\right)\\
       &=\mathtt{EV}_{i}\left(h, \theta_{i}\middle|\tau, \left(\beta'_{i,t}, \beta_{-(i,t)}\right), \left(\pi'_{i,t}, \pi_{-(i,t)}\right);V_{i}\right),
   \end{aligned}
   \]
which implies 
\[
J_{i}\left(h\middle|\tau, \left(\beta'_{i,t}, \beta_{-(i,t)}\right),  \pi  \right)\geq 
J_{i}\left(h\middle|\tau, \left(\beta'_{i,t}, \beta_{-(i,t)}\right), \left(\pi'_{i,t}, \pi_{-(i,t)}\right)  \right).
\]
However, due to (\ref{eq:proof_subperfect_1}), we have
\[
J_{i}\left(h\middle|\tau, \left(\beta'_{i,t}, \beta_{-(i,t)}\right),  \pi  \right)> J_{i}\left(h\middle|\tau, \beta, \pi\right),
\]
which contradicts to (\ref{eq:ppme_refined}).
Thus, we complete the proof of the lemma.
\hfill $\square$

\subsection{Proof of Proposition \ref{prop:opt}}\label{app:prop:opt}

Suppose that $\left<\beta, \pi\right>$ with $V$ is a PPME. 
By the definition of PPME, it is straightforward to show that the constraints (\ref{eq:FE1}), (\ref{eq:FE2}), (\ref{eq:OB}), and (\ref{eq:EQ}) are satisfied.
Hence, $\left<\pi, V\right>$ is a feasible solution to the optimization problem of (\ref{set:PPME}).
By construction, $Z\left(\pi, V\middle| \beta,\tau,C\right)=0$.
From the feasibility, $\left<\beta,\pi\right>$ is a global minimum of the optimization problem of (\ref{set:PPME}).

Conversely, suppose that $\left<\pi, V\right>\in \mathcal{K}\left(\beta\middle| \tau, C\right)$ with $Z\left(\pi, V\middle| \beta,\tau,C\right)=0$.
Then, the constraints (\ref{eq:OB}) and (\ref{eq:EQ}) imply that, for all $i\in\mathcal{N}$, $h\in\mathcal{H}$, $\theta_{i}\in \Theta_{i}$ with $\tau_{i}(\theta_{i}|s, \beta_{i}(h))>0$ where $s\in\mathcal{S}$ with $T(s|h)>0$,
\[
\begin{aligned}
    V_{i}\left(h, \theta\right) \geq \sum_{a} \mathbf{Q}_{i}\left(h, \theta_{i}, a\middle| \beta, \tau; V_{i}\right)\pi\left(a\middle|\theta\right).
\end{aligned}
\]
However $Z\left(\pi, V\middle| \beta,\tau,C\right)=0$.
Then, we obtain, for all $i\in \mathcal{N}$
\[
\begin{aligned}
    V_{i}\left(h, \theta\right) = \sum_{a} \mathbf{Q}_{i}\left(h, \theta_{i}, a\middle| \beta, \tau; V_{i}\right)\pi\left(a\middle|\theta\right).
\end{aligned}
\]
By iteration, we have that $V$ is the unique optimal HT value function profile associated with $\pi$. 
In addition, the constraint (\ref{eq:OB}) implies that given $V$, $\beta$ is a PPME selection policy profile. 
Therefore, the profile $\left<\beta, \pi\right>$ is a PPME.
\hfill $\square$

\subsection{Proof of Proposition \ref{prop:FPA}}\label{app:prop:FPA}

Suppose that $\left<\pi, V\right>\in \mathcal{E}\left(\beta\middle|\tau, C\right)$, i.e., $\left<\beta, \pi, V\right>$ is a global minimum of the optimization problem in (\ref{set:PPME}) with 
$Z\left(\pi, V\middle|\beta, \tau, C\right)=0$.
Then, the constraints (\ref{eq:regular_tau}) and (\ref{eq:feasible_tau}) are trivially satisfied.
Proposition \ref{prop:opt} implies that $\left<\beta, \pi\right>$ is a PPME.
From the construction of $Z(\cdot)$ in (\ref{eq:obj_1}) and the constraint (\ref{eq:EQ}), we have that $\left<\tau,\pi, V\right>$ satisfies (\ref{eq:EQ4}).
According to (\ref{eq:history_value_function_1}), we construct $J$ as $ J_{i}(h) = \sum_{\theta, s}V_{i}\left(h,\theta\middle|\tau,\beta,\pi\right)\tau\left(\theta\middle|s,\beta\left(h\right)\right)T(s|h)$.
Then, $Z^{\mathtt{FPA}}\left(\tau, J, V\middle|\pi, C\right)=0$.
Since $\left<\pi, V\right>$ satisfies (\ref{eq:OB}) given $\tau$, 
\[
J_{i}(h)\geq \sum_{\theta_{-i}, s} V_{i}\left(h, \theta_{i}, \theta_{-i}\right)\tau_{-i}\left(\theta_{-i}\middle|s,\beta_{-i}(h) \right)T(s|h),
\]
for all $\theta_{i}\in\Theta_{i}$ and $h\in\mathcal{H}$, which implies (\ref{eq:OB1}).
From the constraints (\ref{eq:OB1}) and (\ref{eq:EQ4}), we know that for any feasible $\left<\tau', J', V'\right>$, $Z^{\mathtt{FPA}}\left(\tau', J', V'\middle| \pi'\right)\geq0$ where $\pi'$ is the corresponding policy profile.
Therefore, from $Z^{\mathtt{FPA}}\left(\tau, J, V\middle|\pi, C\right)=0$, we conclude that $\left<\tau, J, V\right>$ is a global minimum of the optimization problem in (\ref{set:FPA}).

Conversely, let $\left<\tau, J, V\right>\in \mathcal{E}^{\mathtt{FPA}}\left(\pi, C\right)$ with $Z^{\mathtt{FPA}}(\tau, J, V|\pi)=0$.
Then
\begin{equation}\label{eq:proof_FPA_1}
        J_{i}(h) = \sum_{\theta, s}V_{i}\left(h,\theta\middle|\tau,\beta,\pi\right)\tau\left(\theta\middle|s,\beta\left(h\right)\right)T(s|h).
\end{equation}
The constraint (\ref{eq:EQ4}) directly implies (\ref{eq:EQ}), while the constraint (\ref{eq:OB1}) implies
\begin{equation}\label{eq:proof_FPA_2}
    \begin{aligned}
        J_{i}\left(h\right) &\geq \sum\limits_{s, \theta_{-i},a}\Bigg(\left( \overline{R}_{i}(h, \theta_{i}, a) + J_{i}\left(s, a\right)\right) \\&
        \times\pi(a|\theta_{i},\theta_{-i})\tau_{}\left(\theta_{i},\theta_{-i}|s, \beta(h)\right) T(s|h)\Bigg),
    \end{aligned}
\end{equation}
where the RHS can be written as
\[
\begin{aligned}
    \textup{RHS of } (\ref{eq:proof_FPA_2}) &= \sum_{s, \theta_{-i}, a} \Bigg(\left(\overline{R}_{i}(h, \theta_{i}, a) + \sum_{s}J_{i}\left(s, a\right) \mu_{i}(s|\theta_{i}, h)\right)\\
    &\times\pi(a|\theta)\tau_{-i}\left(\theta_{-i}|s, \beta(h)\right) T(s|h)\Bigg).
\end{aligned}
\]
Construct 
\[
\begin{aligned}
    &\mathbf{Q}_{i}\left(h, \theta_{i}, a\middle|\tau, \beta;V_{i}\right)= \overline{R}_{i}\left(h, \theta_{i}, a\right)\\
    &+ \delta \sum\limits_{s}   \sum_{\theta, s} V_{i}\left(h, \theta\middle|\tau, \beta,\pi\right) \tau\left(\theta\middle|s, \beta\left(h\right)\right)T\left(s\middle|h\right) \mu_{i}\left(s\middle|\theta_{i};\tau\right).
\end{aligned}
\]
Then,
\[
\begin{aligned}
    \textup{RHS of } (\ref{eq:proof_FPA_2}) &=  \sum_{s, \theta_{-i}, a} \Bigg(\left(\mathbf{Q}_{i}\left(h, \theta_{i}, a\middle|\tau, \beta;V_{i}\right)\right)\\
    &\times\pi(a|\theta)\tau_{-i}\left(\theta_{-i}|s, \beta(h)\right) T(s|h)\Bigg).
\end{aligned}
\]
The constraint (\ref{eq:EQ4}) implies
$ 
V_{i}\left(h,\theta\right) =$ $ \sum_{a}\mathbf{Q}_{i}\left(h, \theta_{i}, a\middle|\tau, \beta;V_{i}\right)\pi(a|\theta),
$ 
and thus $Z(\pi, V|\tau) = 0$.
Hence, from (\ref{eq:proof_FPA_1}) and (\ref{eq:proof_FPA_2}), we have
\[
\begin{aligned}
    \sum_{\theta,s}&V_{i}\left(h,\theta\right)\tau(\theta|s,\beta(h))T(s|h) \\
    &\geq \sum_{\theta_{-i},s}V_{i}\left(h,\theta\right)\tau_{-i}(\theta_{-i}|s,\beta_{-i}(h))T(s|h),
\end{aligned}
\]
for all $\theta_{i}\in \Theta_{i}$, which implies (\ref{eq:OB}).
Therefore, $\left<\pi, V\right>\in \mathcal{E}\left(\beta\middle|\tau, C\right)$  with $Z(\pi, V|\tau) = 0$.
\hfill $\square$

%%%%%%%
\subsection{Proof of Theorem \ref{thm:local_admissible}}\label{app:thm:local_admissible}

To prove Theorem \ref{thm:local_admissible}, we show that $\left<\tau, J,V \right>\in\mathcal{E}^{\mathtt{GFPA}}\left(\pi\right)$ with $Z^{\mathtt{GFPA}}\left(\tau, J, V\middle|\pi \right)=0$ if and only if $\left<\tau,\pi\right>$ is locally admissible.

\vspace{0.2cm}
\noindent\textit{Local Admissibility $\Longrightarrow$ PPME} 
\vspace{0.2cm}

Fix any $s\in S$ and $h\in H$.
Suppose that $\left<\tau, \pi\right>$ is locally admissible.
From $\Delta_{i}\left(X^{s}_{i}, \tau_{i}, \bm{f}_{i}\right) = 0$, 
\[
\gradient_{X^{s}_{i}} Z^{\mathtt{LFPA}}_{i}(X^{s}_{i}, \tau_{i}, s, h) - \sum\nolimits_{\theta_{i}\in \Theta_{i}}f[\theta_{i}]\gradient_{X^{s}_{i}}\lambda_{i}(X^{s}_{i}; \theta_{i}, h)=0
\]
Since $\left\{\gradient_{X^{s}_{i}} \lambda_{i}(X^{s}_{i}; \theta_{i}, h)\right\}_{\theta_{i}\in\Theta_{i}}$ is a set of linearly independent vectors for all $X^{s}_{i}$, $i\in\mathcal{N}$, $s\in \mathcal{S}$, $h\in \mathcal{H}$, we have, for all $i\in\mathcal{N}$,
\begin{equation}\label{eq:thm_la_1}
    f\left[\theta_{i}\right] = \tau_{i}\left(\theta_{i}\middle|s, g_{i}, h\right), \textup{ for all } \theta_{i}\in\Theta_{i}.
\end{equation}
In the decomposition $\Theta_{i} = \Theta^{\natural}_{i}\bigcup \left\{ \hat{\theta}_{i}\right\}$, $\hat{\theta}_{i}$ can be fully characterized by $\Theta^{\natural}_{i}$. That is,
\[
\tau_{i}\left(\hat{\theta}_{i}\middle|s, g_{i}, h\right) = 1- \sum\nolimits_{\theta_{i}\in\Theta^{\natural}_{i}}\tau_{i}\left(\theta_{i}\middle| s, g_{i}, h \right).
\]
From $D_{i}\left(X^{s}_{i}, \tau_{i}(\theta_{i}|\cdot), e_{i}, b[\theta_{i}]| s, h\right)=0$,  we have $b[\theta_{i}] - e + \frac{\partial}{\partial \tau_{i}(\theta_{i}|\cdot)} M_{i}(X^{s}_{i}, \tau_{i}(\theta_{i}|\cdot), s, h) = 0$.
Then, $b[\theta_{i}] = e + \lambda_{i}\left(X^{s}_{i}; \hat{\theta}_{i}, h\right) - \lambda_{i}\left(X^{s}_{i}; \theta_{i}, h\right)$ and $e = -\lambda_{i}\left(X^{s}_{i}; \hat{\theta}_{i}, h\right) + \lambda_{i}\left(X^{s}_{i}; \theta_{i}, h\right) + b[\theta_{i}]$.
% %
% \[
% \begin{aligned}
%     &b[\theta_{i}] = e + \lambda_{i}\left(X^{s}_{i}; \hat{\theta}_{i}, h\right) - \lambda_{i}\left(X^{s}_{i}; \theta_{i}, h\right),\\
%     &e = -\lambda_{i}\left(X^{s}_{i}; \hat{\theta}_{i}, h\right) + \lambda_{i}\left(X^{s}_{i}; \theta_{i}, h\right) + b[\theta_{i}].
% \end{aligned}
% \]
% %
From (\ref{eq:thm_la_1}) and $\mathbf{K}(\bm{e}, \bm{b}, \bm{f}; \tau, \bm{\lambda})=0$, we have, for all $\theta_{i}\in \Theta^{\natural}_{i}$,
\[
\begin{aligned}
    &\tau_{i}\left(\theta_{i}\middle|s, g_{i}, h\right)\left(\lambda_{i}\left(X^{s}_{i}; \hat{\theta}_{i}, h\right) - \lambda_{i}\left(X^{s}_{i}; \theta_{i}, h\right) +e  \right) =0\\
    & \tau_{i}\left(\hat{\theta}_{i}\middle|s, g_{i}, h\right)\left(-\lambda_{i}\left(X^{s}_{i}; \hat{\theta}_{i}, h\right) + \lambda_{i}\left(X^{s}_{i}; \theta_{i}, h\right) + b[\theta_{i}]\right) =0,
\end{aligned}
\]
and for all $\theta_{i}\in \Theta_{i}$,
\[
\tau_{i}\left(\theta_{i}\middle|s, g_{i}, h\right)\lambda_{i}\left(X^{s}_{i}; \theta_{i}, h \right),
\]
which implies
\begin{equation}\label{eq:thm_la_2}
    \begin{aligned}
        & b[\theta_{i}] = -\lambda_{i}\left(X^{s}_{i}; \theta_{i}, h \right), \forall \theta_{i}\in \Theta^{\natural}_{i},\\
        & e= -\lambda_{i}\left(X^{s}_{i}; \hat{\theta}_{i}, h \right).
    \end{aligned}
\end{equation}
Therefore, $\bm{F}\left(\bm{X}^{s}, \tau, \bm{e}, \bm{b}, \bm{f}\right)=0$ and $\bm{K}\left(\bm{e}, \bm{b}, \bm{f} ;  \tau,  \bm{\lambda}\right) = 0$ imply $Z^{\mathtt{LFPA}}_{i}(X^{s}_{i}, \tau_{i}, s, h)=0$, leading to $Z^{\mathtt{GFPA}}\left(\tau, J, V\middle|\pi \right)=0$.
In addition, $\pi_{i}(a_{i}|\theta_{i})\gamma_{i}\left(J_{i}, V_{i}, \pi_{-i}| \tau_{i}, \theta_{i}, a_{i}, h\right) = 0$ implies that $Z(\pi, V|\tau)=0$.
From Proposition \ref{prop:opt}, $\left<\tau, \pi\right>$ with $V$ constitutes a PPME.
Then, from Proposition \ref{prop:FPA}, we have $\left<\tau, J, V\right>\in\mathcal{E}^{\mathtt{GFPA}}\left(\pi, C \right)$ with $Z^{\mathtt{GFPA}}\left(\tau, J, V \middle|\pi\right)=0$.

\vspace{0.2cm}
\noindent\textit{PPME $\Longrightarrow$ Local Admissibility} 
\vspace{0.2cm}

Suppose that $\left<\tau, \pi\right>$ is a PPME.
Hence, Proposition \ref{prop:FPA} implies $\left<\tau, J, V\right>\in\mathcal{E}^{\mathtt{GFPA}}\left(\pi\right)$ with $Z^{\mathtt{GFPA}}\left(\tau, J, V \middle|\pi\right)=0$.
Then, it holds for every $i\in\mathcal{N}$ that
\[
J_{i}(h) \geq \sum\nolimits_{\theta_{-i}, s} V_{i}(h, \theta_{i}, \theta_{-i})\tau(\theta_{i}, \theta_{-i}|s, g, h)T_{s}(s|h),
\]
for all $i\in\mathcal{N}$, $h\in\mathcal{H}$, $\theta_{i}\in\Theta_{i}$.
which implies that $\lambda_{i}(X^{s}_{i}; \theta_{i}, h)\geq 0$.
Since $Z^{\mathtt{GFPA}}\left(\tau, J, V \middle|\pi\right)=0$, we have
\[
J_{i}(h) = \sum\nolimits_{\theta, s} V_{i}(h, \theta)\tau(\theta|s, g, h)T_{s}(s|h).
\]
Then, from the definition of $Z^{\mathtt{LFPA}}_{i}$ in (\ref{eq:misalign_def}), $Z^{\mathtt{LFPA}}_{i}(X^{s}_{i}, \tau_{i}, s, h)=0$.
Since $\lambda_{i}(X^{s}_{i}; \theta_{i}, h)\geq 0$ for all $\theta_{i}\in\Theta_{i}$, we have
\[
\tau_{i}(\theta_{i}|s, g, h) \lambda_{i}(X^{s}_{i}; \theta_{i}, h)=0.
\]
By constructing $f[\theta_{i}]$ according to (\ref{eq:thm_la_1}) and $b[\theta_{i}]$ and $e$ according to (\ref{eq:thm_la_2}), respectively, we can show that there exist Lagrange multipliers such that the conditions in $\mathcal{R}\left(s,h\right)$ are satisfied.

From Proposition \ref{prop:opt}, $\left<\pi, V\right>\in \mathcal{E}\left(\beta\middle|\tau\right)$  with $Z(\pi, V|\tau) = 0$.
Hence, we have
\[
\mathtt{EV}_{i}\left(h, \theta_{i}|\tau_{i}, \pi, V_{i}\right) -\mathbb{E}^{\mu_{i}}_{\pi_{-i}}\left[ Q_{i}(h, \theta_{i}, a_{i}, \tilde{a}_{-i}|\tau; J_{i})\Big|h, \theta_{i} \right]\geq0.
\]
However, $Z(\pi, V|\tau) = 0$. Then, it holds that
\[
\mathtt{EV}_{i}\left(h, \theta_{i}|\tau_{i}, \pi, V_{i}\right) -\mathbb{E}^{\mu_{i}}_{\pi_{-i}}\left[ Q_{i}(h, \theta_{i}, a_{i}, \tilde{a}_{-i}|\tau; J_{i})\Big|h, \theta_{i} \right] = 0.
\]
Therefore, we have $\pi_{i}(a_{i}|\theta_{i})\gamma_{i}\left(J_{i}, V_{i}, \pi_{-i}| \tau_{i}, \theta_{i}, a_{i}, h\right) = 0$, for all $i\in N, a_{i}\in A_{i}, \theta_{i}\in\Theta_{i}, h\in\mathcal{H}, (\ref{eq:FE1}), (\ref{eq:FE2})$.
Thus, we conclude that $\left<\tau,\pi \right>$ is locally admissible.
\hfill $\square$

%%%%%%%
\subsection{Proof of Theorem \ref{thm:local_admissible}}\label{app:thm:local_admissible}

To prove Theorem \ref{thm:local_admissible}, we show that $\left<\tau, J,V \right>\in\mathcal{E}^{\mathtt{GFPA}}\left(\pi\right)$ with $Z^{\mathtt{GFPA}}\left(\tau, J, V\middle|\pi \right)=0$ if and only if $\left<\tau,\pi\right>$ is locally admissible.

\vspace{0.2cm}
\noindent\textit{Local Admissibility $\Longrightarrow$ PPME} 
\vspace{0.2cm}

Fix any $s\in S$ and $h\in H$.
Suppose that $\left<\tau, \pi\right>$ is locally admissible.
From $\Delta_{i}\left(X^{s}_{i}, \tau_{i}, \bm{f}_{i}\right) = 0$, 
\[
\gradient_{X^{s}_{i}} Z^{\mathtt{LFPA}}_{i}(X^{s}_{i}, \tau_{i}, s, h) - \sum\nolimits_{\theta_{i}\in \Theta_{i}}f[\theta_{i}]\gradient_{X^{s}_{i}}\lambda_{i}(X^{s}_{i}; \theta_{i}, h)=0
\]
Since $\left\{\gradient_{X^{s}_{i}} \lambda_{i}(X^{s}_{i}; \theta_{i}, h)\right\}_{\theta_{i}\in\Theta_{i}}$ is a set of linearly independent vectors for all $X^{s}_{i}$, $i\in\mathcal{N}$, $s\in \mathcal{S}$, $h\in \mathcal{H}$, we have, for all $i\in\mathcal{N}$,
\begin{equation}\label{eq:thm_la_1}
    f\left[\theta_{i}\right] = \tau_{i}\left(\theta_{i}\middle|s, g_{i}, h\right), \textup{ for all } \theta_{i}\in\Theta_{i}.
\end{equation}
In the decomposition $\Theta_{i} = \Theta^{\natural}_{i}\bigcup \left\{ \hat{\theta}_{i}\right\}$, $\hat{\theta}_{i}$ can be fully characterized by $\Theta^{\natural}_{i}$. That is,
\[
\tau_{i}\left(\hat{\theta}_{i}\middle|s, g_{i}, h\right) = 1- \sum\nolimits_{\theta_{i}\in\Theta^{\natural}_{i}}\tau_{i}\left(\theta_{i}\middle| s, g_{i}, h \right).
\]
From $D_{i}\left(X^{s}_{i}, \tau_{i}(\theta_{i}|\cdot), e_{i}, b[\theta_{i}]| s, h\right)=0$,  we have $b[\theta_{i}] - e + \frac{\partial}{\partial \tau_{i}(\theta_{i}|\cdot)} M_{i}(X^{s}_{i}, \tau_{i}(\theta_{i}|\cdot), s, h) = 0$.
Then, $b[\theta_{i}] = e + \lambda_{i}\left(X^{s}_{i}; \hat{\theta}_{i}, h\right) - \lambda_{i}\left(X^{s}_{i}; \theta_{i}, h\right)$ and $e = -\lambda_{i}\left(X^{s}_{i}; \hat{\theta}_{i}, h\right) + \lambda_{i}\left(X^{s}_{i}; \theta_{i}, h\right) + b[\theta_{i}]$.
% %
% \[
% \begin{aligned}
%     &b[\theta_{i}] = e + \lambda_{i}\left(X^{s}_{i}; \hat{\theta}_{i}, h\right) - \lambda_{i}\left(X^{s}_{i}; \theta_{i}, h\right),\\
%     &e = -\lambda_{i}\left(X^{s}_{i}; \hat{\theta}_{i}, h\right) + \lambda_{i}\left(X^{s}_{i}; \theta_{i}, h\right) + b[\theta_{i}].
% \end{aligned}
% \]
% %
From (\ref{eq:thm_la_1}) and $\mathbf{K}(\bm{e}, \bm{b}, \bm{f}; \tau, \bm{\lambda})=0$, we have, for all $\theta_{i}\in \Theta^{\natural}_{i}$,
\[
\begin{aligned}
    &\tau_{i}\left(\theta_{i}\middle|s, g_{i}, h\right)\left(\lambda_{i}\left(X^{s}_{i}; \hat{\theta}_{i}, h\right) - \lambda_{i}\left(X^{s}_{i}; \theta_{i}, h\right) +e  \right) =0\\
    & \tau_{i}\left(\hat{\theta}_{i}\middle|s, g_{i}, h\right)\left(-\lambda_{i}\left(X^{s}_{i}; \hat{\theta}_{i}, h\right) + \lambda_{i}\left(X^{s}_{i}; \theta_{i}, h\right) + b[\theta_{i}]\right) =0,
\end{aligned}
\]
and for all $\theta_{i}\in \Theta_{i}$,
\[
\tau_{i}\left(\theta_{i}\middle|s, g_{i}, h\right)\lambda_{i}\left(X^{s}_{i}; \theta_{i}, h \right),
\]
which implies
\begin{equation}\label{eq:thm_la_2}
    \begin{aligned}
        & b[\theta_{i}] = -\lambda_{i}\left(X^{s}_{i}; \theta_{i}, h \right), \forall \theta_{i}\in \Theta^{\natural}_{i},\\
        & e= -\lambda_{i}\left(X^{s}_{i}; \hat{\theta}_{i}, h \right).
    \end{aligned}
\end{equation}
Therefore, $\bm{F}\left(\bm{X}^{s}, \tau, \bm{e}, \bm{b}, \bm{f}\right)=0$ and $\bm{K}\left(\bm{e}, \bm{b}, \bm{f} ;  \tau,  \bm{\lambda}\right) = 0$ imply $Z^{\mathtt{LFPA}}_{i}(X^{s}_{i}, \tau_{i}, s, h)=0$, leading to $Z^{\mathtt{GFPA}}\left(\tau, J, V\middle|\pi \right)=0$.
In addition, $\pi_{i}(a_{i}|\theta_{i})\gamma_{i}\left(J_{i}, V_{i}, \pi_{-i}| \tau_{i}, \theta_{i}, a_{i}, h\right) = 0$ implies that $Z(\pi, V|\tau)=0$.
From Proposition \ref{prop:opt}, $\left<\tau, \pi\right>$ with $V$ constitutes a PPME.
Then, from Proposition \ref{prop:FPA}, we have $\left<\tau, J, V\right>\in\mathcal{E}^{\mathtt{GFPA}}\left(\pi, C \right)$ with $Z^{\mathtt{GFPA}}\left(\tau, J, V \middle|\pi\right)=0$.

\vspace{0.2cm}
\noindent\textit{PPME $\Longrightarrow$ Local Admissibility} 
\vspace{0.2cm}

Suppose that $\left<\tau, \pi\right>$ is a PPME.
Hence, Proposition \ref{prop:FPA} implies $\left<\tau, J, V\right>\in\mathcal{E}^{\mathtt{GFPA}}\left(\pi\right)$ with $Z^{\mathtt{GFPA}}\left(\tau, J, V \middle|\pi\right)=0$.
Then, it holds for every $i\in\mathcal{N}$ that
\[
J_{i}(h) \geq \sum\nolimits_{\theta_{-i}, s} V_{i}(h, \theta_{i}, \theta_{-i})\tau(\theta_{i}, \theta_{-i}|s, g, h)T_{s}(s|h),
\]
for all $i\in\mathcal{N}$, $h\in\mathcal{H}$, $\theta_{i}\in\Theta_{i}$.
which implies that $\lambda_{i}(X^{s}_{i}; \theta_{i}, h)\geq 0$.
Since $Z^{\mathtt{GFPA}}\left(\tau, J, V \middle|\pi\right)=0$, we have
\[
J_{i}(h) = \sum\nolimits_{\theta, s} V_{i}(h, \theta)\tau(\theta|s, g, h)T_{s}(s|h).
\]
Then, from the definition of $Z^{\mathtt{LFPA}}_{i}$ in (\ref{eq:misalign_def}), $Z^{\mathtt{LFPA}}_{i}(X^{s}_{i}, \tau_{i}, s, h)=0$.
Since $\lambda_{i}(X^{s}_{i}; \theta_{i}, h)\geq 0$ for all $\theta_{i}\in\Theta_{i}$, we have
\[
\tau_{i}(\theta_{i}|s, g, h) \lambda_{i}(X^{s}_{i}; \theta_{i}, h)=0.
\]
By constructing $f[\theta_{i}]$ according to (\ref{eq:thm_la_1}) and $b[\theta_{i}]$ and $e$ according to (\ref{eq:thm_la_2}), respectively, we can show that there exist Lagrange multipliers such that the conditions in $\mathcal{R}\left(s,h\right)$ are satisfied.

From Proposition \ref{prop:opt}, $\left<\pi, V\right>\in \mathcal{E}\left(\beta\middle|\tau\right)$  with $Z(\pi, V|\tau) = 0$.
Hence, we have
\[
\mathtt{EV}_{i}\left(h, \theta_{i}|\tau_{i}, \pi, V_{i}\right) -\mathbb{E}^{\mu_{i}}_{\pi_{-i}}\left[ Q_{i}(h, \theta_{i}, a_{i}, \tilde{a}_{-i}|\tau; J_{i})\Big|h, \theta_{i} \right]\geq0.
\]
However, $Z(\pi, V|\tau) = 0$. Then, it holds that
\[
\mathtt{EV}_{i}\left(h, \theta_{i}|\tau_{i}, \pi, V_{i}\right) -\mathbb{E}^{\mu_{i}}_{\pi_{-i}}\left[ Q_{i}(h, \theta_{i}, a_{i}, \tilde{a}_{-i}|\tau; J_{i})\Big|h, \theta_{i} \right] = 0.
\]
Therefore, we have $\pi_{i}(a_{i}|\theta_{i})\gamma_{i}\left(J_{i}, V_{i}, \pi_{-i}| \tau_{i}, \theta_{i}, a_{i}, h\right) = 0$, for all $i\in N, a_{i}\in A_{i}, \theta_{i}\in\Theta_{i}, h\in\mathcal{H}, (\ref{eq:FE1}), (\ref{eq:FE2})$.
Thus, we conclude that $\left<\tau,\pi \right>$ is locally admissible.
\hfill $\square$

\subsection{Proof of Theorem \ref{thm:equivalence_PI_PPME}}

     Consider $\left<\tau, \pi\right>$ that is a PPME of a game $\mathtt{B}^{\Gamma}$ with any cognition cost profile $C\in \mathcal{F}$.
     Hence, the profile $\left<\tau, \pi\right>$ and the base game model $\mathtt{B}$ induces the probability measures $P\left[\tau,\pi\right]$, $P\left[\tau,\pi\middle|h\right]$, and $P\left[\tau,\pi\middle|h, \theta_{i}\right]$.
     With abuse of notation, we use $f(\cdot)$ to denote the marginal mass or density function corresponding to $P\left[\tau,\pi\right]$; e.g., $f(a_{t}|h_{t})$, $f(c_{i,t}|h_{t})$, $f(a_{i,t}|h_{t}, \theta_{i,t})$.
     Given the PPME $\left<\tau,\pi\right>$ of a game $\mathtt{B}^{\Gamma}$ with feasible cognition profile $\Gamma$, consider a profile $\left< \xi,\pi^{*}\right>$ and a \texttt{SBA} cost profile $C^{*}$ that satisfy the following, for all $i\in\mathcal{N}$, $t\geq 1$, 
         \[
         \begin{aligned}
             & \pi^{*}_{i,t}(a_{i,t}|s_{t}) \equiv f(a_{i,t}|s_{t}) \textup{ and } C^{*}_{i,t}(s_{t}, a_{i,t})\equiv \int_{\mathcal{C}} c_{i,t} f(c_{i,t}|s_{t}, a_{i,t}).
         \end{aligned}
         \]
         %
         % Then, $\left<\xi, \pi^{*} \right>$ is a PI-PPME of $\mathtt{B}^{\Gamma\left[\xi\right]}$ with feasible cognition profile $\Gamma\left[\xi\right]$ where each cognition cost is given by $C^{*}_{i,t}(s_{t}, a_{i,t})$. 
%
     In the PPME $\left<\tau,\pi\right>$ of the game $\mathbb{B}^{\Gamma}$, the H value function for any $h_{t}\in \mathcal{H}$ can be given by, 
     \begin{equation}
              \begin{aligned}
         &J_{i,t}\left(h_{t}\middle| \tau, \pi\right)=\mathbb{E}^{\tau}_{\pi}\left[R_{i}\left(\tilde{s}_{i,t}, \tilde{a}_{t}\right) + \tilde{c}_{i,t} +\delta J_{i,t+1}\left(\tilde{s}_{i,t}, \tilde{a}_{t}\middle| \tau, \pi\right)  \middle| h_{t}\right]\\
         &=\mathbb{E}^{\tau}_{\pi}\left[ R_{i}\left(\tilde{s}_{i,t}, \tilde{a}_{t}\right)\middle| h_{t}\right] + \mathbb{E}^{\tau}_{\pi}\left[ \tilde{c}_{i,t}\middle| h_{t}\right] +\delta\mathbb{E}^{\tau}_{\pi}\left[J_{i,t+1}\left(\tilde{s}_{t}, \tilde{a}_{t}\middle| \pi, \tau \right) \middle| h_{t}\right]. 
         %
         % \\
         % &+ \mathbb{E}^{\tau}_{\pi}\left[ \mathbb{E}^{\tau}_{\pi}\left[ \sum_{k=t+1}^{K} \delta^{k-t-1}\left( R_{i}\left(\tilde{s}_{i,k}, \tilde{a}_{k}\right) + \tilde{c}_{i,k} \right) +\delta^{K-t} J_{i,T+1}\left(\tilde{s}_{i,T}, \tilde{a}_{T} \right) \middle| \tilde{s}_{i,t}, \tilde{a}_{t}\right]\middle|h_{t}\right].
     \end{aligned}
     \end{equation}
     % \textcolor{red}{for any $K>t$},
     % \begin{equation}
     %          \begin{aligned}
     %     &J_{i,t}\left(h_{t}\middle| \tau, \pi\right)=\mathbb{E}^{\tau}_{\pi}\left[R_{i}\left(\tilde{s}_{i,t}, \tilde{a}_{t}\right) + \tilde{c}_{i,t} +\delta J_{i,t+1}\left(\tilde{s}_{i,t}, \tilde{a}_{t}\middle| \tau, \pi\right)  \middle| h_{t}\right]\\
     %     &=\mathbb{E}^{\tau}_{\pi}\left[ R_{i}\left(\tilde{s}_{i,t}, \tilde{a}_{t}\right)\middle| h_{t}\right] + \mathbb{E}^{\tau}_{\pi}\left[ \tilde{c}_{i,t}\middle| h_{t}\right] \\
     %     &+ \mathbb{E}^{\tau}_{\pi}\left[ \mathbb{E}^{\tau}_{\pi}\left[ \sum_{k=t+1}^{K} \delta^{k-t-1}\left( R_{i}\left(\tilde{s}_{i,k}, \tilde{a}_{k}\right) + \tilde{c}_{i,k} \right) +\delta^{K-t} J_{i,T+1}\left(\tilde{s}_{i,T}, \tilde{a}_{T} \right) \middle| \tilde{s}_{i,t}, \tilde{a}_{t}\right]\middle|h_{t}\right].
     % \end{aligned}
     % \end{equation}
     %
     Given the profile $\left<\xi, \pi^{*} \right>$, the H value function for any $h_{t}\in\mathcal{H}$ can be represented in terms of the EP-HSA value function as follows:
     \begin{equation}
         \begin{aligned}
             &J_{i,t}\left(h_{t} \middle|\xi, \pi^{*}\right) = \mathbb{E}^{\xi}_{\pi^{*}}\left[ W_{i,t}\left(\tilde{s}_{t}, \tilde{a}_{t}\middle|\xi,\pi^{*}\right)\middle|h_{t}\right] \\
             &=\sum_{s_{t}, a_{t}} W_{i,t}\left(s_{t}, a_{t}\middle|\xi,\pi^{*}\right) \pi^{*}_{t}(a_{t}|s_{t})T(s_{t}|h_{t})\\
             &=\sum_{s_{t}, a_{t}} R_{i}(s_{t}, a_{t})\pi^{*}_{t}(a_{t}|s_{t})T(s_{t}|h_{t}) \\
             &+ \sum_{s_{t}, a_{t}} C^{*}_{i,t}(s_{t}, a_{t})\pi^{*}_{t}(a_{t}|s_{t})T(s_{t}|h_{t})\\
             &+ \delta\sum_{s_{t}, a_{t}} J_{i,t+1}\left(s_{t}, a_{t} \middle|\xi, \pi^{*}\right)\pi^{*}_{t}(a_{t}|s_{t})T(s_{t}|h_{t})\\
             &= \sum_{s_{t}, a_{t}} R_{i}(s_{t}, a_{t})f(a_{t}|s_{t})T(s_{t}|h_{t}) \\
             &+ \sum_{s_{t}, a_{t}}\int_{\mathcal{C}}c_{i,t} f(c_{i,t}|s_{t}, a_{t})  f(a_{t}|s_{t})T(s_{t}|h_{t})\\
             &+ \delta\sum_{s_{t}, a_{t}} J_{i,t+1}\left(s_{t}, a_{t} \middle|\xi, \pi^{*}\right)f(a_{t}|s_{t})T(s_{t}|h_{t}).
         \end{aligned}
     \end{equation}
Given $P\left[\tau,\pi\right]$, it holds that $f(a_{t}|s_{t})T(s_{t}|h_{t}) = f(s_{t}, a_{t}|h_{t})$. Hence,
\[
\begin{aligned}
    &J_{i,t}\left(h_{t} \middle|\xi, \pi^{*}\right) = \mathbb{E}^{\xi}_{\pi^{*}}\left[ W_{i,t}\left(\tilde{s}_{t}, \tilde{a}_{t}\middle|\xi,\pi^{*}\right)\middle|h_{t}\right]\\
    &=\mathbb{E}^{\tau}_{\pi}\left[R_{i}\left(\tilde{s}_{t}, \tilde{a}_{t}\right) + \tilde{c}_{i,t}+ \delta\sum_{s_{t}, a_{t}} J_{i,t+1}\left(\tilde{s}_{t}, \tilde{a}_{t} \middle|\xi, \pi^{*}\right)\middle| h_{t}\right]\\
    &=\mathbb{E}^{\tau}_{\pi}\left[W_{i,t}\left(\tilde{s}_{t}, \tilde{a}_{t} \middle|\xi, \pi^{*}\right) \middle| h_{t}\right]= J_{i,t}\left(h_{t}\middle|\tau,\pi\right).
\end{aligned}
\]

Next, we prove that the profile $\left<\xi, \pi^{*}\right>$ with $C^{*}$ is indeed a PI-PPME.
We proceed with the proof by showing a contradiction.
Let $\hat{\xi}=\xi\circ\tau_{i,t}\equiv\left(\xi_{1}, \dots, \xi_{t-1}, (\hat{\tau}_{i,t},\xi_{-i,t}), \xi_{t+1},\dots\right)$ denote a profile that is the same as $\xi$ except for agent $i$'s period-$t$ choice $\tau_{i,t}$.
Define $\hat{\pi}=\pi^{*}\circ \hat{\pi}_{i,t}$ in the same way such that $\hat{\pi}\in \Pi\left[\hat{\xi}; \mathtt{B}^{\Gamma\left[\hat{\xi}\right]}\right]$.
In addition, the cognition cost $C^{*}$ remains the same.
Given any history $h_{t}\in\mathcal{H}$, the the profile $\left<\hat{\xi}, \hat{\pi} \right>$ induces the H value function as
\[
\begin{aligned}
    &J_{i,t}\left(h_{t}\middle| \hat{\xi}, \hat{\pi} \right)\\
    &=  \sum_{s_{t}, a_{t}, \theta_{i,t}} R_{i}(s_{t}, a_{t})f(a_{t}|s_{t},\theta_{i,t})\hat{\tau}_{i,t}(\theta_{i,t}|s_{t},h_{t})T(s_{t}|h_{t}) \\
             &+ \sum_{s_{t}, a_{t}, \theta_{i,t}}\int_{\mathcal{C}}c_{i,t} f(c_{i,t}|s_{t}, a_{t})  f(a_{t}|s_{t}, \theta_{i,t})\hat{\tau}_{i,t}(\theta_{i,t}|s_{t}, h_{t}) T(s_{t}|h_{t})\\
             &+ \delta\sum_{s_{t}, a_{t}, \theta_{i,t}} J_{i,t+1}\left(s_{t}, a_{t} \middle|\xi, \pi^{*}\right)f(a_{t}|s_{t},\theta_{i,t})\hat{\tau}_{i,t}(\theta_{i,t}|s_{t}, h_{t}) T(s_{t}|h_{t}).
\end{aligned}
\]
Here, $f(a_{t}|s_{t},\theta_{i,t})\hat{\tau}_{i,t}(\theta_{i,t}|s_{t}, h_{t}) T(s_{t}|h_{t})\neq f(s_{t},a_{t},\theta_{t}|h_{t})$ because $f(\cdot)$ is corresponding to $P\left[\tau, \pi\right]$.
If $\left<\xi, \pi^{*}\right>$ with $C^{*}$ is not a PI-PPME, then there must exist a history $h_{t}$ and a profile $\left<\hat{\xi},\hat{\pi} \right>$ such that $J_{i,t}(h_{t}|\hat{\xi}, \hat{\tau})>  J_{i,t}(h_{t}|\xi, \pi^{*}) = J_{i,t}(h_{t}|\tau,\pi)$, which implies that $\left<\tau,\pi\right>$ can be strictly improved by unilateral deviation $(\hat{\xi}_{i,t}, \hat{\pi}_{i,t})$ which contradicts the fact that $\left<\tau,\pi\right>$ is PPME.
\hfill $\square$

%
%================================
\end{document}